\documentclass[11pt,a4paper]{article}
\usepackage{amsfonts}
\usepackage{amsmath}           
\usepackage{amssymb}
\usepackage{amstext}
\usepackage[affil-it]{authblk}
\usepackage[backref=false,backend=bibtex,style=numeric-comp,sorting=none,maxcitenames=1,block=none,giveninits,url=false]{biblatex}
\usepackage{bm}                
\usepackage{booktabs}
\usepackage{color}
\usepackage{enumerate}
\usepackage{fancyhdr}
\usepackage{float}
\usepackage{graphicx}
\usepackage[space]{grffile}    
\usepackage[top=2.0cm, right=2.0cm, left=2.0cm, bottom=2.0cm]{geometry}
\usepackage[colorlinks=false, pdfborder={0 0 0}, breaklinks]{hyperref}
\usepackage[utf8]{inputenc}
\usepackage[switch]{lineno}    
\usepackage{longtable}
\usepackage{marvosym}
\usepackage{multicol}
\usepackage{multirow}
\usepackage{setspace}
\usepackage[separate-uncertainty,multi-part-units = single, allow-number-unit-breaks]{siunitx}
\usepackage{subfig}
\usepackage[nottoc,numbib]{tocbibind} 
\usepackage{tikz-timing}
\usepackage[colorinlistoftodos,prependcaption,textsize=tiny]{todonotes} 
\usepackage{wasysym}

\usepackage{setspace}
\usepackage{url}

\modulolinenumbers[5]

\usepackage{csquotes}

\newcommand*\patchAmsMathEnvironmentForLineno[1]{%
  \expandafter\let\csname old#1\expandafter\endcsname\csname #1\endcsname
  \expandafter\let\csname oldend#1\expandafter\endcsname\csname end#1\endcsname
  \renewenvironment{#1}%
     {\linenomath\csname old#1\endcsname}%
     {\csname oldend#1\endcsname\endlinenomath}}%
\newcommand*\patchBothAmsMathEnvironmentsForLineno[1]{%
  \patchAmsMathEnvironmentForLineno{#1}%
  \patchAmsMathEnvironmentForLineno{#1*}}%
\AtBeginDocument{%
\patchBothAmsMathEnvironmentsForLineno{equation}%
\patchBothAmsMathEnvironmentsForLineno{align}%
\patchBothAmsMathEnvironmentsForLineno{flalign}%
\patchBothAmsMathEnvironmentsForLineno{alignat}%
\patchBothAmsMathEnvironmentsForLineno{gather}%
\patchBothAmsMathEnvironmentsForLineno{multline}%
}

\newcommand{\ar}{$\textrm{Ar}$}

\newcommand{\co}{$\textrm{CO}_{2}$}
\newcommand{\ch}{$\textrm{CH}_{4}$}

\newcommand{\arco}{$\textrm{Ar}\textnormal{-}\textrm{CO}_2$}
\newcommand{\arcois}[1]{$\textrm{Ar}\textnormal{-}\textrm{CO}_2$ (#1)}

\newcommand{\arch}{$\textrm{Ar}\textnormal{-}\textrm{CH}_4$}
\newcommand{\archis}[1]{$\textrm{Ar}\textnormal{-}\textrm{CH}_4$ (#1)}
\newcommand{\neconis}[1]{$\textrm{Ne}\textnormal{-}\textrm{CO}_2\textnormal{-}\textrm{N}_2$ (#1)}

\newcommand{\fe}[1]{$^{#1}\textrm{Fe}$}

\newcommand{\figref}[1]{Figure~\ref{#1}}    
\newcommand{\figrefbra}[1]{Fig.~\ref{#1}}   
\newcommand{\Figref}[1]{Figure~\ref{#1}}    
\newcommand{\tabref}[1]{Table~\ref{#1}}      
\newcommand{\Tabref}[1]{Table~\ref{#1}}     
\newcommand{\secref}[1]{Section~\ref{#1}}   
\newcommand{\secrefbra}[1]{Sec.~\ref{#1}}   
\newcommand{\Secref}[1]{Section~\ref{#1}}   
\usepackage[T1]{fontenc}
\usepackage[UKenglish]{babel}

\DeclareSIUnit\bar{bar}

\title{\vspace{-5ex}First operation of an ALICE OROC operated in high pressure \arco{} and \arch{}}
\date{\vspace{-10ex}}

\author[a]{A.~Ritchie-Yates\thanks{Corresponding author: \texttt{harrison.ritchie-yates.2013@live.rhul.ac.uk (A.R.-Y.)}}}
\author[a]{A.~Deisting\thanks{Now at: Johannes Gutenberg-Universität Mainz, Institut für Physik, 55128 Mainz (DE)}}
\author[b]{G.~Barker} 
\author[b]{S.~Boyd}
\author[c]{D.~Brailsford}
\author[a]{Z.~Chen-Wishart}
\author[d]{L.~Cremonesi}
\author[e]{P.~Dunne}
\author[a]{J.~Eeles}
\author[e]{P.~Hamilton}
\author[a]{A.C.~Kaboth}
\author[e]{N.~Khan}
\author[e]{A.~Klustová}
\author[a]{J.~Monroe}
\author[c]{J.~Nowak}
\author[d]{P.~Singh}
\author[d]{A.V.~Waldron}
\author[a]{J.~Walding}
\author[e]{L.~Warsame}
\author[e]{M.O.~Wascko}
\author[e]{I.~Xiotidis}

 \affil[a]{Royal Holloway, University of London, United Kingdom}
 \affil[b]{University of Warwick, United Kingdom}
 \affil[c]{Lancaster University, Bailrigg, Lancaster, United Kingdom}
 \affil[d]{Queen Mary, University of London, United Kingdom}
 \affil[e]{Imperial College London, United Kingdom}

\def\FigList{}
\newcommand\AddFigToList[1]{\xdef\FigList{\FigList #1 ,}}

\LetLtxMacro{\OldIncludegraphics}{\includegraphics}
\renewcommand{\includegraphics}[2][]{%
    \AddFigToList{#2}%
    \OldIncludegraphics[#1]{#2}%
}

\newcommand*{\ShowListOfFigures}{%
    \typeout{Figures included were}%
    \foreach \x in \FigList {%
        \typeout{ \x}
    }%
    \typeout{End of list of figures}%
}
\AtEndDocument{\ShowListOfFigures}

\begin{document}

\maketitle

\begin{abstract}
New neutrino-nucleus interaction cross-section measurements are required to improve nuclear models sufficiently for future experiments like the Deep Underground Neutrino Experiment or Hyper-Kamiokande to meet their sensitivity goals. A time projection chamber (TPC) filled with a high-pressure gas is a promising detector to characterise the high-intensity neutrino sources planned for future long-baseline experiments. A gas-filled TPC is ideal for measuring low-energy particles as they travel much further in gas than solid or liquid neutrino detectors. Using a high-pressure gas increases the target density, resulting in more neutrino interactions. This paper will examine the suitability of multiwire proportional chambers (MWPCs) taken from the ALICE TPC to be used as the readout chambers of a high-pressure gas TPC. Such a design could be suitable as the near detector for future long-baseline experiments. These chambers were previously operated at atmospheric pressure. We tested one such MWPC at up to almost 5 bar absolute (barA) with the UK high-pressure test stand at Royal Holloway, University of London.\\
{\indent}This paper reports the successful operation of an ALICE TPC outer readout chamber (OROC) at pressures up to \SI{4.8}{bar} absolute with \arch{} mixtures with a \ch{} content between \SI{2.8}{\%} and \SI{5.0}{\%}, and so far up to \SI{4}{bar} absolute with \arcois{90-10}. We measured the charge gain of this OROC using signals induced by an \fe{55} source. The largest gain achieved at \SI{4.8}{bar} was $(64\pm2)\cdot10^{3}$ at stable conditions with an anode wire voltage of \SI{2990}{\volt} in \archis{95.9-4.1}. In \arcois{90-10} a gain of $(4.2\pm0.1)\cdot10^{3}$ was observed at an anode voltage of \SI{2975}{\volt} at \SI{4}{barA} gas pressure. Based on all our gain measurements, we extrapolate that, at the \SI{10}{barA} pressure necessary to fit 1 tonne of gas into the ALICE TPC volume, a gain of 5000 in \arcois{90-10} (10000 in \arch{} with $\sim\!\SI{4}{\%}$ \ch{} content) may be achieved with an OROC anode voltage of \SI{4.2}{\kilo\volt} ($\sim\!\SI{3.1}{\kilo\volt}$).
These voltages are above the maximum anode voltage at which the OROC has been tested (\SI{3}{\kilo\volt}), so further study of the upper limit of $V_a$ or further optimisation of the gas mixes may be necessary for future operation of OROCs at pressures up to \SI{10}{barA}.
\end{abstract}

\section{Introduction}
\label{sec:intro}

A time projection chamber (TPC) filled with a high pressure gas is a promising detector to characterise interactions from high-intensity neutrino sources, \textit{e.g.} the neutrino beams of future long-baseline neutrino experiments such as the Deep Underground Neutrino Experiment (DUNE) and Hyper-Kamiokande (HK). Using this kind of detector as part of a near detector complex provides substantial robustness to oscillation analyses, for example, as studied in the DUNE near detector (ND) conceptual design report \cite{NDCDR2021}.\\
{\indent}High resolution measurements of neutrino-nucleus interaction cross-sections are required as an input parameter to improve the cross section models used for neutrino oscillation analyses. In the near future, the uncertainties related to these models will be the main contribution to the systematic uncertainties of oscillation analyses so they need to be reduced in order for future neutrino oscillation experiments (like DUNE and HK) to reach their sensitivity goals \cite{ALVAREZRUSO20181}. Current experiments like T2K and NO$\nu$A also suffer from mismodelling of neutrino-nucleus interactions. Neutrino interactions with nuclei are measured by their final state particles, which are ejected from the nucleus after an interaction; measuring all (charged) final state particles requires measuring protons and pions with a low kinetic energy of a few \SI{10}{\mega\electronvolt} \cite{NDCDR2021}. A gas filled TPC is ideal for this task. A gas TPC can grant coverage over the full solid angle and, since particles will travel further in a gas medium than they will through a liquid, a gas TPC can measure lower momentum particle tracks compared to a liquid filled TPC. Since neutrinos interact very rarely, even with the MW-scale beams of future neutrino experiments it would be necessary to use a high-pressure gas to increase the target density and allow more scattering events to be recorded.\\
{\indent}The ALICE TPC \cite{ALME2010316} has been recently upgraded with new readout chambers \cite{Adolfsson_2021}, allowing for continuous readout. Consequently, the previously used multiwire proportional chambers (MWPCs) have become available and we will study in this paper how they could be used at a future long-baseline neutrino experiment's ND. These chambers are the appropriate size for a future long-baseline detector, and provide a useful platform for research and development of high pressure TPC (HPTPC) readout technology and test stands. As the ALICE TPC is operated at atmospheric pressure, these MWPCs need to be tested at higher pressures. To achieve the millions of neutrino interactions needed to constrain neutrino cross-sections, approximately a tonne of gas would be needed. Achieving this mass in the volume of ALICE would require a pressure of \SI{10}{bar} absolute (\si{barA}).\\
{\indent}This paper reports on gas gain measurements with an ALICE outer readout chamber (OROC) at the UK high pressure test stand operated at Royal Holloway, University of London (RHUL). In the remainder of this section we will introduce the ALICE TPC and the design of the OROC (\secrefbra{sec:intro:subsec:ALICE}). The UK high pressure platform will be presented in \secref{sec:intro:subsec:RHULHP}. \Secref{sec:setup} describes how these components form the full test set-up and the measurement procedure; and in \secref{sec:analysis} the data analysis is detailed. Results are given in \secref{sec:results}, showing the operation of the OROC at pressures up to \SI{4.8}{barA} with different gas mixtures with argon predominance (\arch{} and \arco{}). We conclude this paper with \secref{sec:summary}.

\subsection{ALICE TPC multiwire proportional chambers}
\label{sec:intro:subsec:ALICE}

\begin{figure}
\centering
\includegraphics[width=0.95\columnwidth]{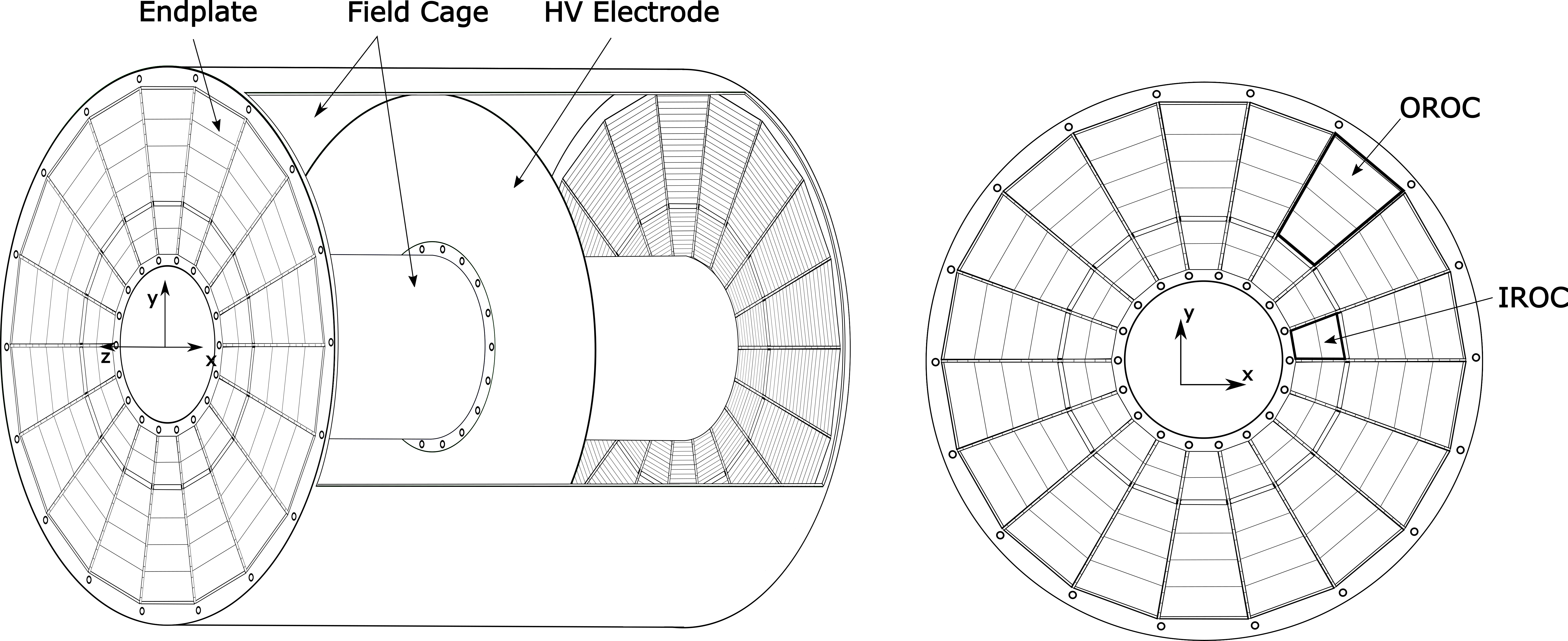}
\caption{\label{sec:intro:subsec:DUNE:fig:ALICE_TPC}Schematic of the ALICE TPC. The TPC features a central drift electrode. Readout chambers are arranged in two endplates; each endplate has 18 sectors of one OROC and one IROC, for a total of 36 readout chambers per endplate. Further tracking detectors and the beam pipe are enclosed in an inner cylinder.}
\end{figure}

ALICE \cite{ALICE2008,ALICE:2014sbx} -- short for \textit{A Large Ion Collider Experiment} -- is an experiment dedicated to the study of the quark gluon plasma in heavy-ion collisions at CERN's LHC. The ALICE TPC \cite{ALME2010316} is the experiment's main tracking and particle identification detector and is optimised to measure the trajectories of charged particles and their specific energy loss in the high multiplicity environment of $\text{Pb}$-$\text{Pb}$ collisions. The original TPC has been designed to track up to 8000 charged particles per rapidity unit per event at a trigger rate of \SI{300}{Hz}. It has so far been operated at atmospheric pressure with \neconis{90-10-5} -- here and throughout the paper the numbers in brackets denote the mixing ratio by volume -- and \arcois{88-12}.\\
{\indent}A sketch of the ALICE TPC is shown in \figref{sec:intro:subsec:DUNE:fig:ALICE_TPC}. It follows a cylindrical design of \SI{5}{\meter} length and \SI{5.16}{\meter} diameter, with a cylindrical cut out of $\SI{157.6}{\centi\meter}$ diameter at the centre. The ROCs sit on the end-plates facing the drift cathode at the centre of the TPC. There are 18 inner readout chambers (IROCs) and OROCs per end-plate. In LHC run 3, ALICE will take data at a rate of \SI{50}{\kilo\hertz} in $\text{Pb}$-$\text{Pb}$ collisions and it will examine every event recorded at that rate. For this reason, the ALICE TPC was upgraded with new readout chambers allowing for continuous readout of the TPC \cite{Adolfsson_2021}. This has made all ALICE TPC MWPCs available to be used by a different experiment, so they could be re-purposed for use in a high-pressure TPC.\\
\begin{figure}
\centering
\subfloat[]{\label{sec:intro:subsec:ALICE:fig:OROC_Geometry:sides}\includegraphics[height=0.25\textheight]{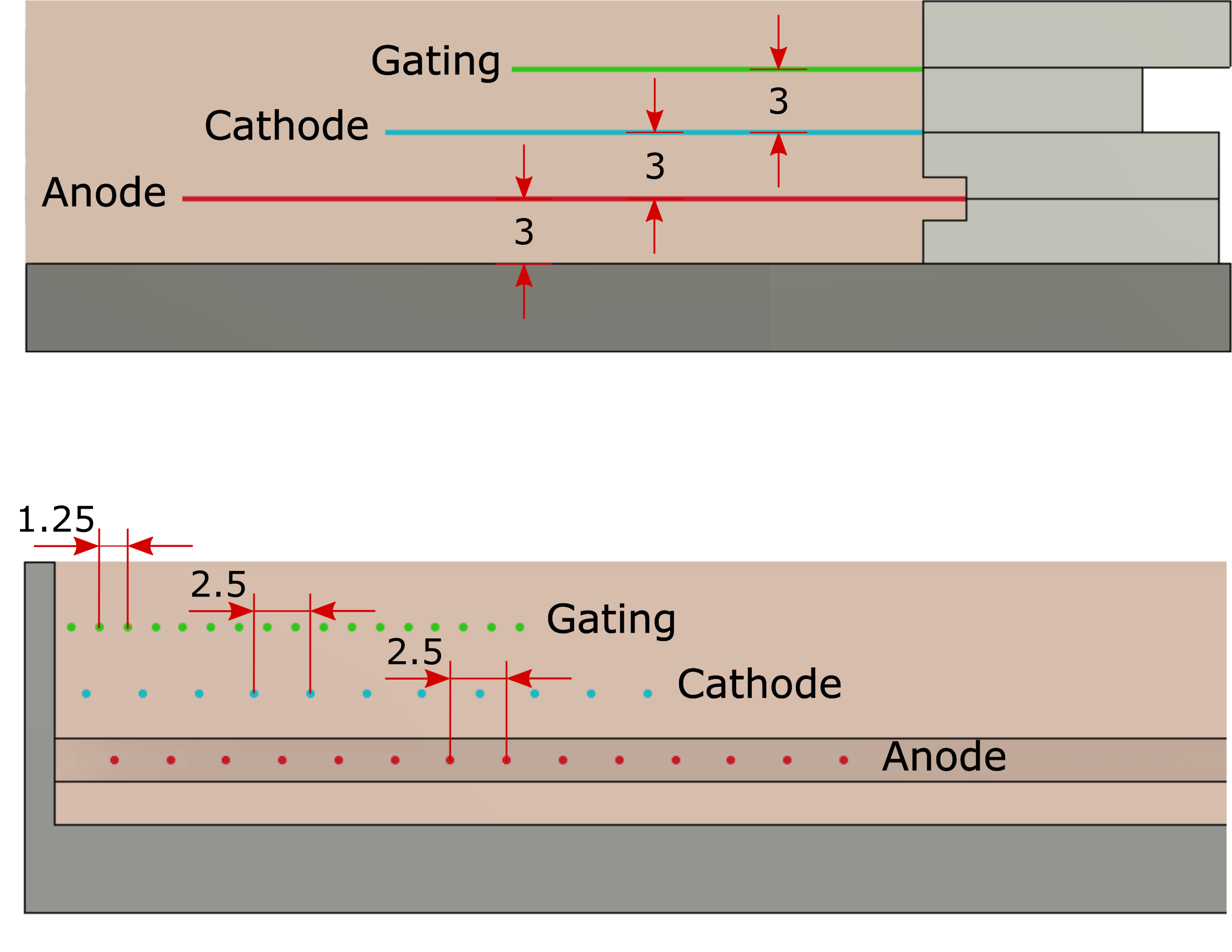}}
\subfloat[]{\label{sec:intro:subsec:ALICE:fig:OROC_Geometry:3d}\includegraphics[height=0.25\textheight]{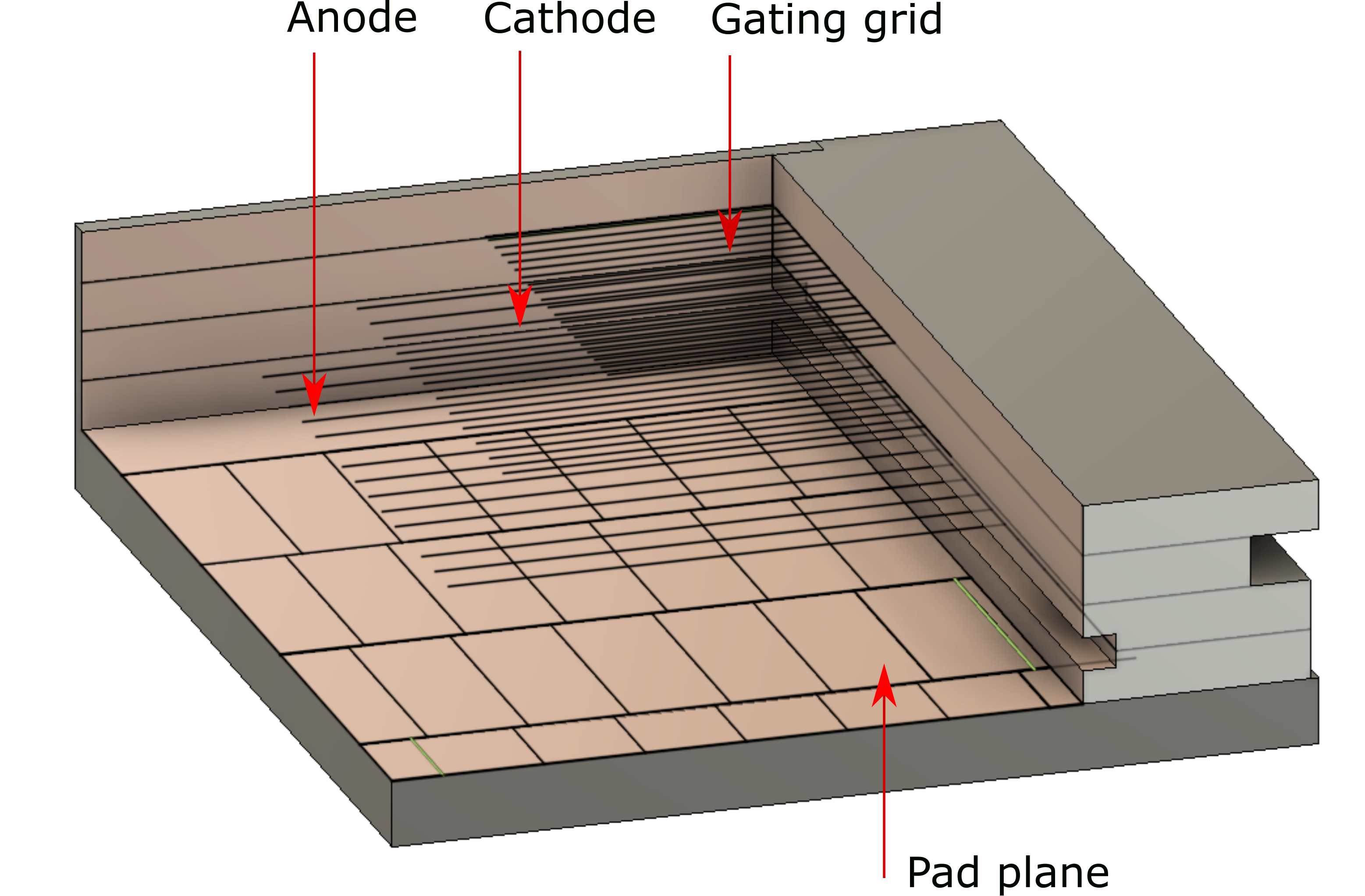}}
\caption{\label{sec:intro:subsec:ALICE:fig:OROC_Geometry}Geometry of the OROC wire- and pad plane layout. \protect\subref{sec:intro:subsec:ALICE:fig:OROC_Geometry:sides} Cross section through the ROC perpendicular (top) and parallel (bottom) to parallel sides of the ROC's trapezoid shape (see \figrefbra{sec:intro:subsec:ALICE:fig:OROC_Diagram}). \protect\subref{sec:intro:subsec:ALICE:fig:OROC_Geometry:3d} A 3D view of one of the ROC's corners.}
\end{figure}
\begin{figure}
\centering
\subfloat[]{\label{sec:intro:subsec:ALICE:fig:OROC_Diagram:pads}\includegraphics[height=0.25\textheight,trim = 0 0 2525 0,clip=true]{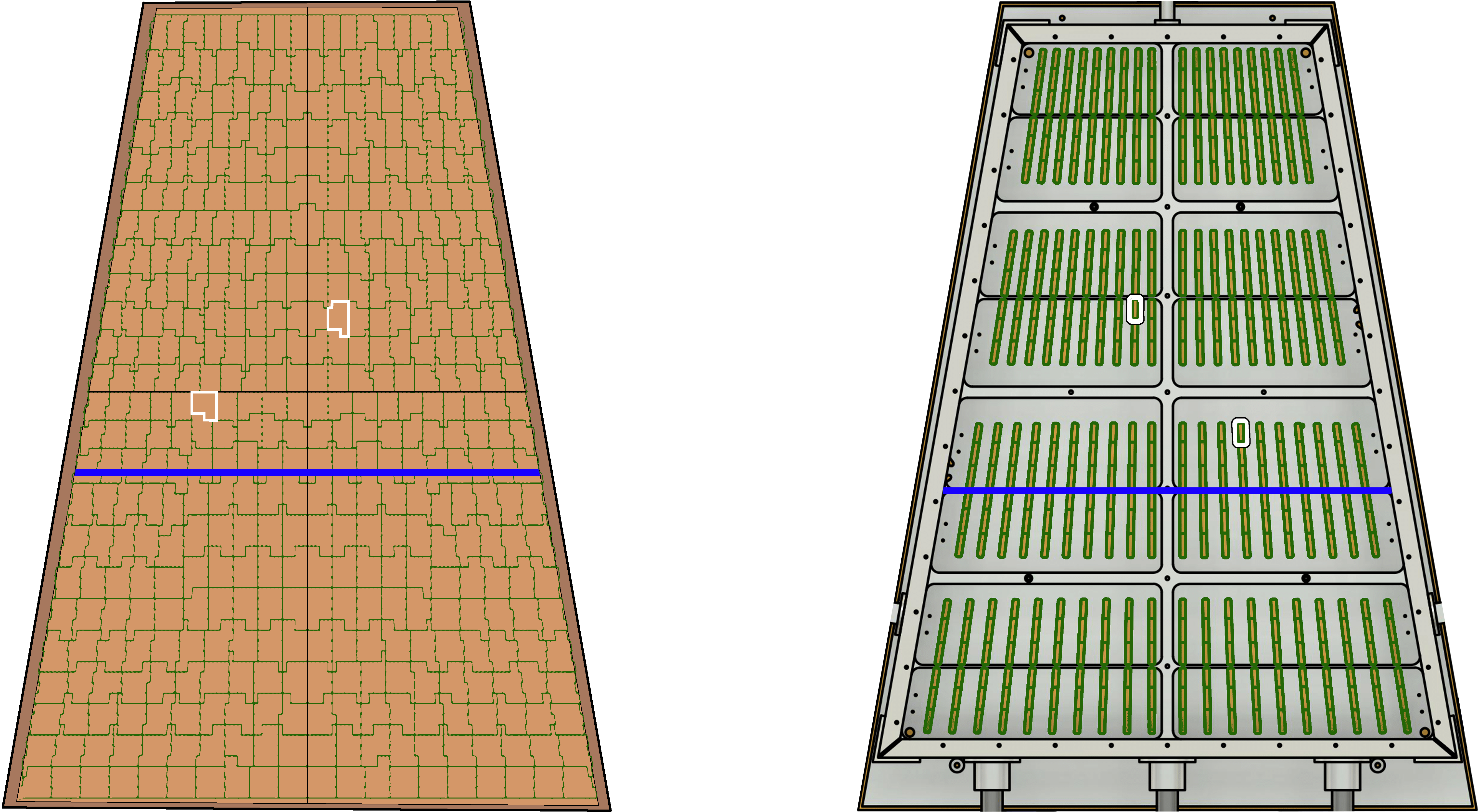}}
\subfloat[]{\label{sec:intro:subsec:ALICE:fig:OROC_Diagram:back}\includegraphics[height=0.25\textheight,trim = 2525 0 0 0,clip=true]{./figures/OROC_Pad_Plane_New7thickblueline}}
\caption{\label{sec:intro:subsec:ALICE:fig:OROC_Diagram}Diagrams of \protect\subref{sec:intro:subsec:ALICE:fig:OROC_Diagram:pads} the pad-plane side of the OROC, showing the wires and pad plane and \protect\subref{sec:intro:subsec:ALICE:fig:OROC_Diagram:back} the connector side of the OROC showing the sockets used for readout and grounding of pads.
The pad plane is made up of 96 rows of two different pad sizes, the boundary between the upper 64 rows of $6\times\SI{10}{\milli\meter\squared}$ pads and the lower 32 rows of larger, $6\times\SI{15}{\milli\meter\squared}$ pads is indicated by the blue line in each diagram.
The pad plane is segmented into groups of 21 or 22 pads, which can be read out from the sockets on the back of the OROC (see \figref{sec:setup:fig:shortingcard}.) The groups and their corresponding sockets have been outlined in green and the two pad groups used for readout for the studies in this paper are outlined in white in each diagram.
}
\end{figure}
{\indent}In \figref{sec:intro:subsec:ALICE:fig:OROC_Geometry} the wire geometry of an OROC is shown. There are three wire planes: the anode wire grid, the cathode wire grid and the gating grid, where the latter separates the TPC drift volume from the actual ROC. The trapezoidal active area (\figrefbra{sec:intro:subsec:ALICE:fig:OROC_Diagram}) of an OROC measures \SI{45.8}{\centi\meter} and \SI{86}{\centi\meter} on the small and on the long end, respectively, defining also the location of the shortest (\SI{43.4}{\centi\meter}) and longest (\SI{83.6}{\centi\meter}) wire length. Wires run in parallel to the parallel sides of the trapezium. The length of the chamber perpendicular to the wires is \SI{114}{\centi\meter}. 
The spacing between each plane of wires is \SI{3}{\milli\meter} and the anode wires sit \SI{3}{\milli\meter} above the pad plane. 
The gating grid wires are \SI{75}{\micro\meter} diameter copper beryllium wires with a pitch of \SI{1.25}{\milli\meter}. An offset voltage of $+\Delta V_{\text{gg}}$ or $-\Delta V_{\text{gg}}$ can be applied on every other wire in addition to the constant gating grid wire voltage $V_{\text{gg}}$, thus creating an electric field configuration in which electrons from the drift volume and ions from the inside of the ROC end up on the gating grid wires and do not cross the grid. The cathode wire grid is also made from copper beryllium wires of \SI{75}{\micro\meter} diameter. Their pitch is \SI{2.5}{\milli\meter}, staggered with respect to the gating grid wires. The particular OROC we used for the studies in this paper was never part of the ALICE TPC, but was used for tests by the ALICE collaboration. As such it has a feed-through, which allows the cathode wire grid to be biased or read out. In the ALICE TPC the cathode wires were held at ground potential. The anode wires are \SI{20}{\micro\meter} in diameter. Like the cathode wires, their pitch is \SI{2.5}{\milli\meter} and they are staggered with respect to both the cathode and gating wire grid. The first two and last two anode wires, called edge wires, are \SI{75}{\micro\meter} in diameter.
The anode wires are designed for high voltage (HV) and so are made of gold-plated tungsten, which has a higher strength than the copper beryllium used for the other wires. The upper voltage limit which the anode wires can maintain without sustaining damage has not been determined, but the OROC has been operated with anode voltages up to \SI{3000}{\volt}.\\ 
{\indent}Ionization electrons are produced in the drift region by charged particles or photons and drift to the amplification region in the ROC under the influence of the electric field produced by the TPC's HV electrode and the gating grid wires. In the ROC these electrons undergo avalanche charge multiplication around the anode wires, inducing charge on the pads. The pad plane is made up of 9984 pads of two different pad sizes. The size of the outer 4032 pads (32 pad rows) is $6\times\SI{15}{\milli\meter\squared}$, the size of the inner 5952 pads (64 pad rows) is $6\times\SI{10}{\milli\meter\squared}$. Outer and inner refers to the pad rows closer to the longer and shorter end of the OROC, respectively. \figref{sec:intro:subsec:ALICE:fig:OROC_Diagram} indicates the border between the rows of different pad sizes. The pads are grouped in groups of 21 or 22 pads and are connected to sockets on the back of the OROC. 
Diagrams of the pads, groups, and sockets are shown in \figref{sec:intro:subsec:ALICE:fig:OROC_Diagram}. In the ALICE TPC, the position of the original ionisation electrons in the drift volume can be determined from the pad signals, which determine the location in $x$ and $y$. The third coordinate is determined by the arrival time of the charge at the ROC, projecting back to a $t_0$ given by a trigger.\\
{\indent}The design of the similar, albeit much smaller IROC design can be found in Ref. \cite{ALME2010316}, together with additional information on the ROCs and the ALICE TPC.

\subsection{UK high pressure platform}
\label{sec:intro:subsec:RHULHP}

\begin{figure}
\centering
\includegraphics[height=0.25\textheight,trim = 0 00 0 00,clip=true]{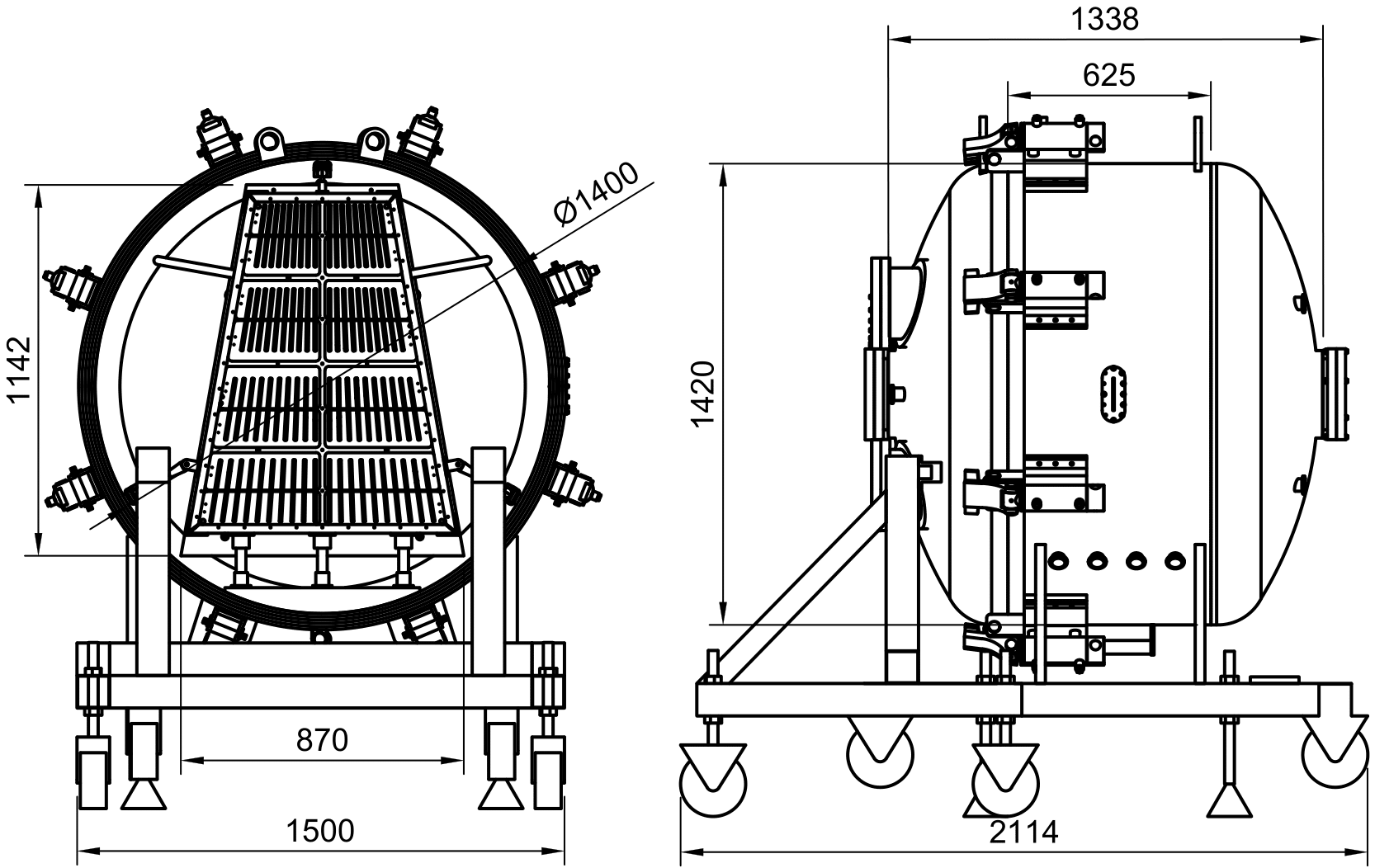}
\caption{\label{sec:setup:fig:VesselDiagram} Schematic drawings of the HPTPC pressure vessel, shown are the front (left), and side (right) views. The front view (cutaway) also shows the OROC support frame mounted on the vessel rails.}
\end{figure}

The UK high pressure test stand operated at RHUL allows for the testing of TPC technology in the pressure range from sub-atmospheric pressure up to \SI{4}{barG}, where the unit \si{barG} is used to indicate gauge pressure. It features a high pressure vessel (\textit{cf}. \figrefbra{sec:setup:fig:VesselDiagram}) with an inner diameter of \SI{140}{\centi\meter}, which is to our knowledge the only vessel which can be used to test ROCs as large as an OROC at pressures higher than atmospheric pressure. A detailed description of the test stand can be found in Ref. \cite{instruments5020022}. The cylindrical vessel with domed ends is made from stainless steel (type 304L) and has a total volume of \SI{1472}{\deci\meter\cubed}. One of the ends can be removed to access the inside of the vessel. A system of three rails allows a TPC with a drift length of more than \SI{50}{\centi\meter} length to be mounted, depending on the width of the TPC's readout structure. The vessel has various feed-throughs to attach HV and signal lines, and to supply the vessel with gas or evacuate it. The vessel at RHUL is connected to a gas system which allows for evacuation to $\SI{1E-6}{barA}$, and gas mixing by partial pressures up to $\SI{5}{barA}$ from 4 inputs.

\section{Experimental set-up and measurement procedure}
\label{sec:setup}

\begin{figure}
\centering
\subfloat[]{\label{sec:setup:fig:TPC_Diagram:side}\includegraphics[height=0.25\textheight]{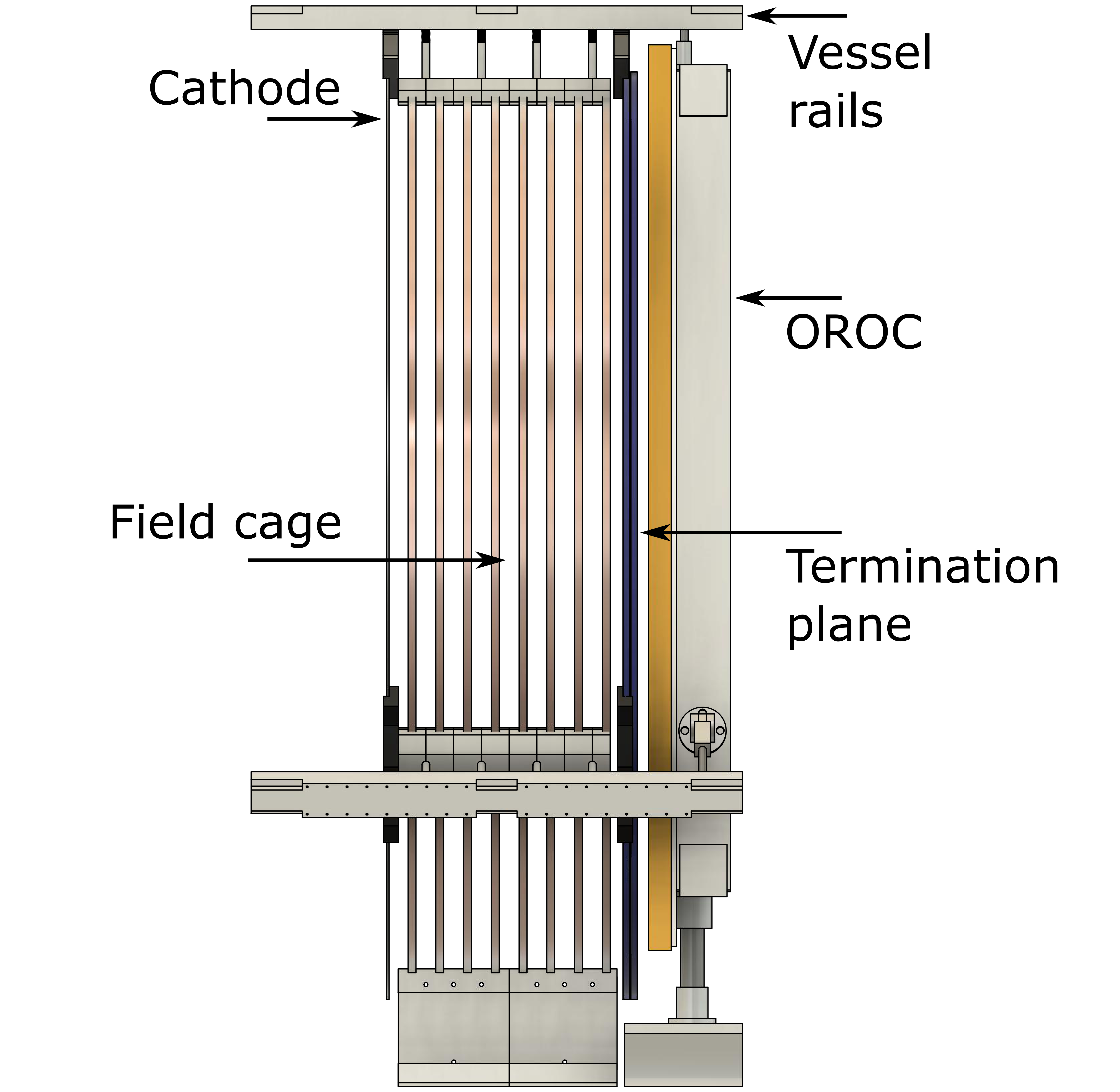}}
\subfloat[]{\label{sec:setup:fig:TPC_Diagram:back}\includegraphics[height=0.25\textheight]{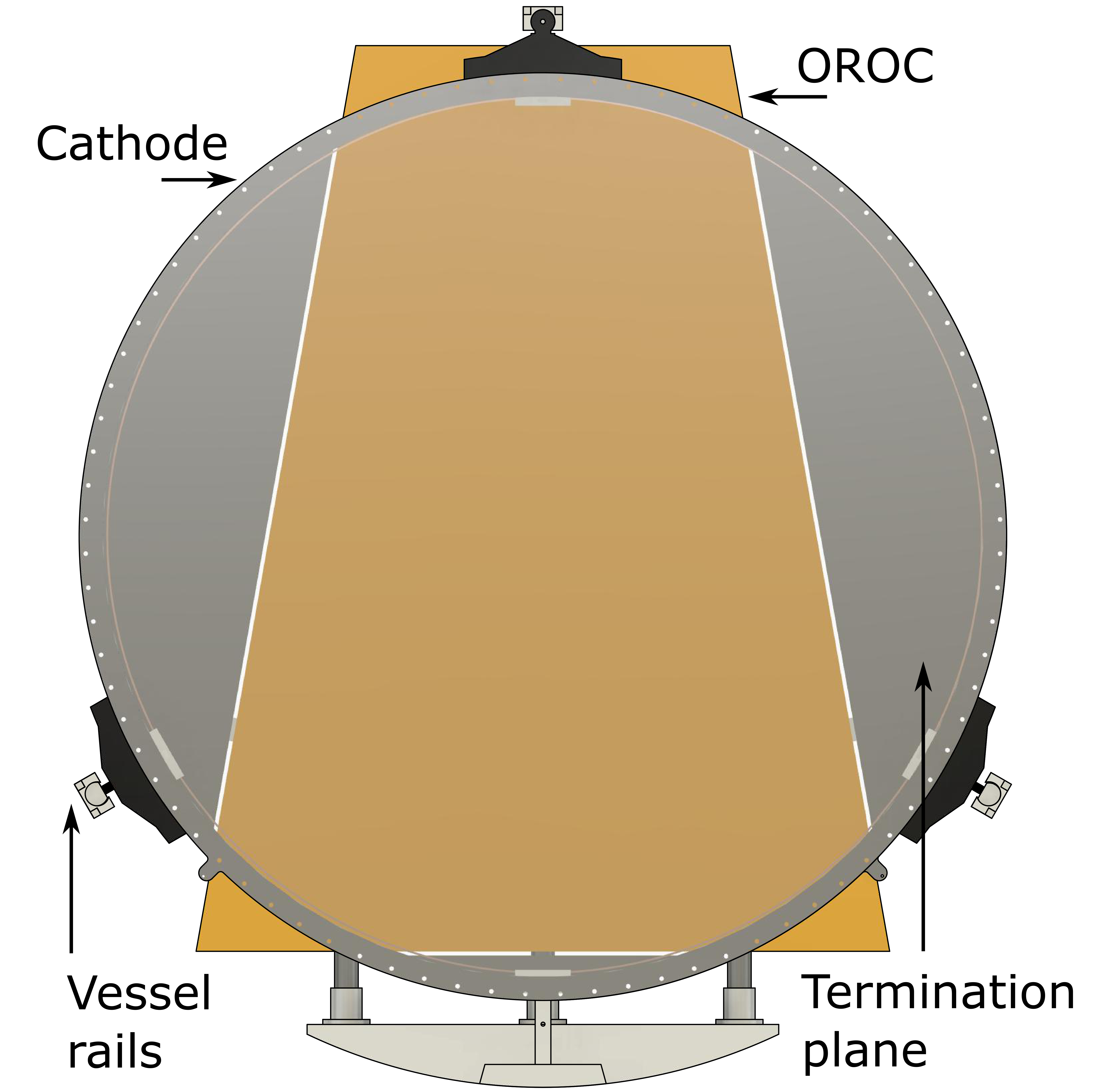}}
\caption{\label{sec:setup:fig:TPC_Diagram}Three dimensional CAD rendering of the TPC, \protect\subref{sec:setup:fig:TPC_Diagram:side} side view and \protect\subref{sec:setup:fig:TPC_Diagram:back} view from the back of the vessel, showing the drift cathode, field cage, terminator plane, and OROC support frame, as well as the rails on which the TPC is mounted in the pressure vessel. Note that the field cage is shown here with 12 rings, the field cage was reduced to 8 rings for the measurements in this paper.}
\end{figure}

\begin{figure}
\centering
\subfloat[]{\label{sec:setup:fig:TPC_photo:front}\includegraphics[height=0.25\textheight]{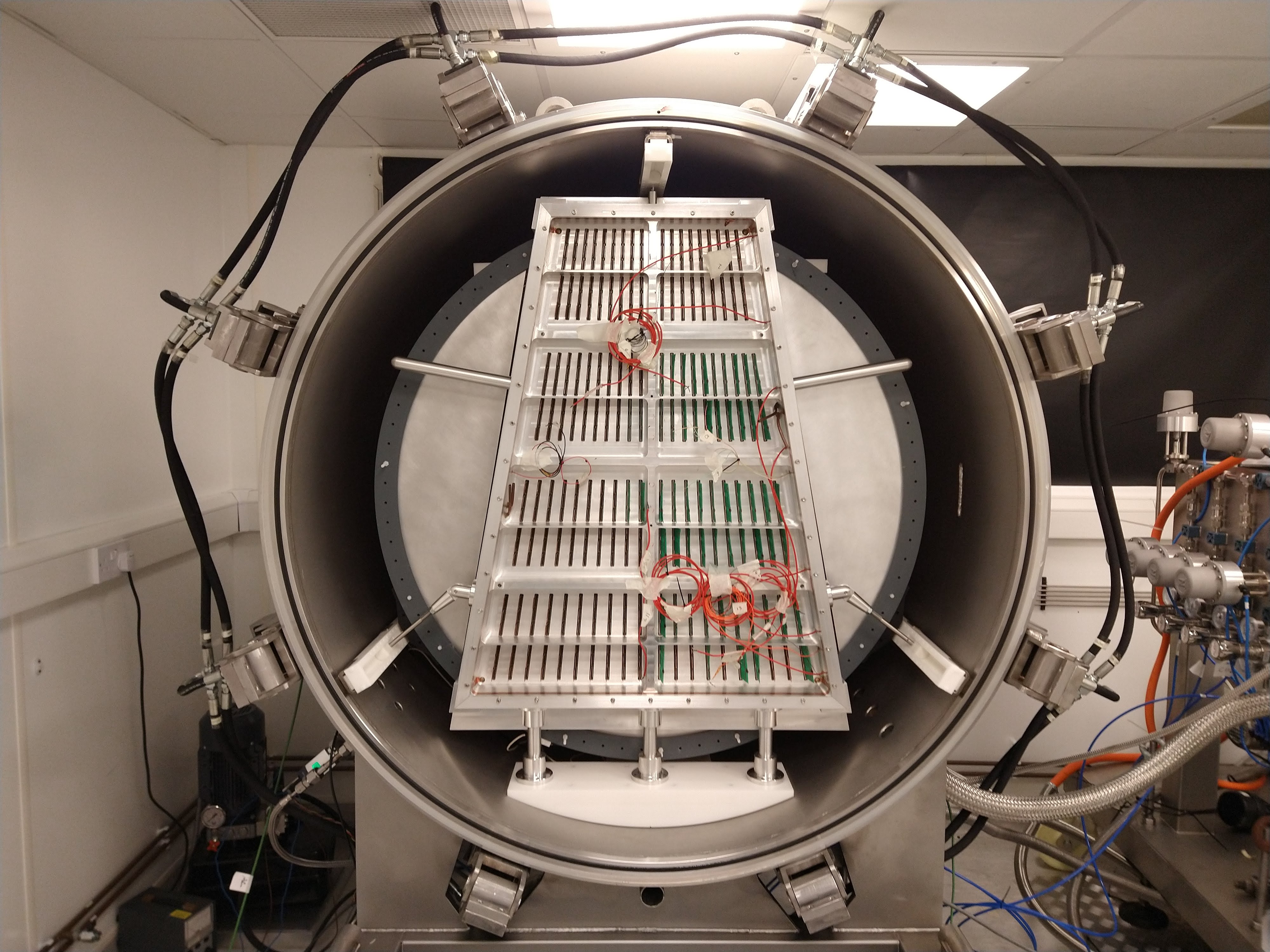}}
\subfloat[]{\label{sec:setup:fig:TPC_photo:34}\includegraphics[height=0.25\textheight]{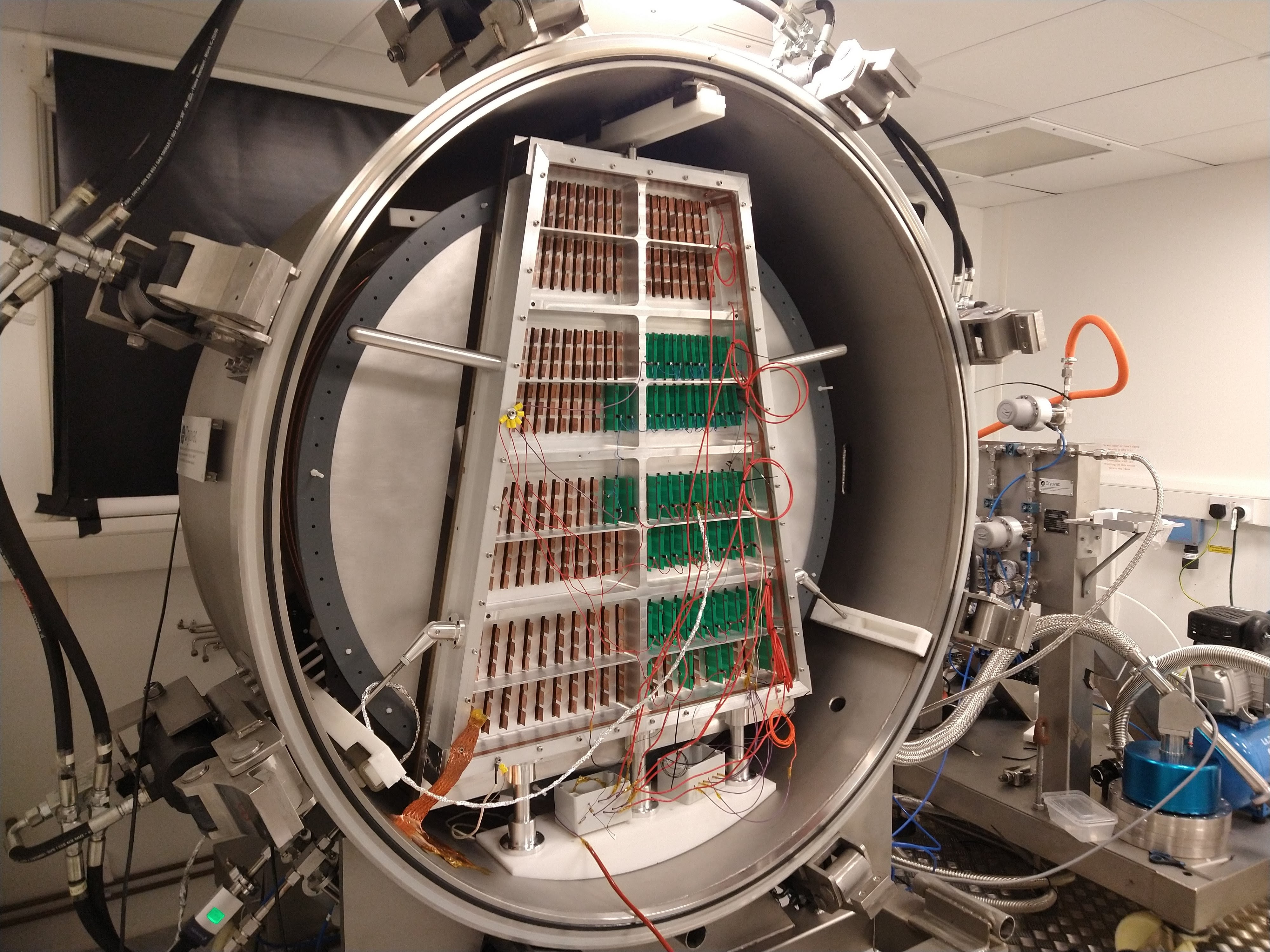}}
\caption{\label{sec:setup:fig:TPC_photo}Photos of the setup at RHUL. The TPC is mounted inside the pressure vessel on three white Delrin rails with the OROC supported by an aluminium frame. The gas system can also be partially seen to the right of both photos. Also visible in \protect\subref{sec:setup:fig:TPC_photo:34} are the HV connections for the anode and gating grid, and the shorting cards used for grounding and readout (see \figref{sec:setup:fig:shortingcard}).}
\end{figure}

For the studies detailed in this paper, the TPC used is mounted inside the high pressure vessel at RHUL on three Delrin rails, \SI{600}{\milli\meter} in length. The drift volume in this set-up is created by the same drift cathode and field cage as described in Ref. \cite{instruments5020022}. The drift cathode is a steel mesh ring with a radius of \SI{1120}{\milli\meter} made from \SI{25}{lpi} (lines per inch) mesh of \SI{27}{\micro\meter} diameter wires. The field cage is made of 8 copper rings with an inner diameter of \SI{1110}{\milli\meter}.
We define a coordinate system such that the pad plane of the OROC is in the $xy$ plane. 
The coordinate along the remaining axis (the TPC symmetry axis) is then $z$.
Each field cage ring has a width of \SI{10}{\milli\meter} in $z$ and the distance between each ring in $z$ is \SI{25}{\milli\meter}, yielding a total field cage length of \SI{255}{\milli\meter}. 
These field-shaping rings are connected by a series of \SI{3}{\mega\ohm} resistors to reduce the potential between each ring. Resistors of the same resistance are used to connect the first ring to the drift electrode and the final ring to ground.\\
{\indent}The shape of the amplification region (\textit{i.e.} the OROC) is trapezoidal but the drift region is cylindrical. To account for this transition, three plates are mounted in a plane between the drift region and amplification region to cover the area of the drift region not covered by the OROC, thus terminating the field lines there. This termination plane is steel with \SI{1}{\milli\meter} thickness and it is also mounted on the same Delrin rails supporting the TPC. A diagram of the TPC is shown in \figref{sec:setup:fig:TPC_Diagram} and a photo of the set-up in \figref{sec:setup:fig:TPC_photo}.\\
{\indent}The total drift distance from the drift electrode to the OROC anode wires is \SI{336}{\milli\meter}. For the data presented in \secref{sec:analysis} the TPC was operated with a drift cathode voltage of \SI{-16}{\kilo\volt} and hence a drift field of \SI{-476}{\volt\per\centi\meter}. The gating grid voltage $V_{\text{gg}}$ was adjusted to the equipotential at the position of the gating grid, \textit{i.e.} \SI{-143}{\volt} for the drift cathode voltage mentioned before. 

\subsection{Charge readout hardware}

\begin{figure}
\centering
\includegraphics[width = 0.45\columnwidth]{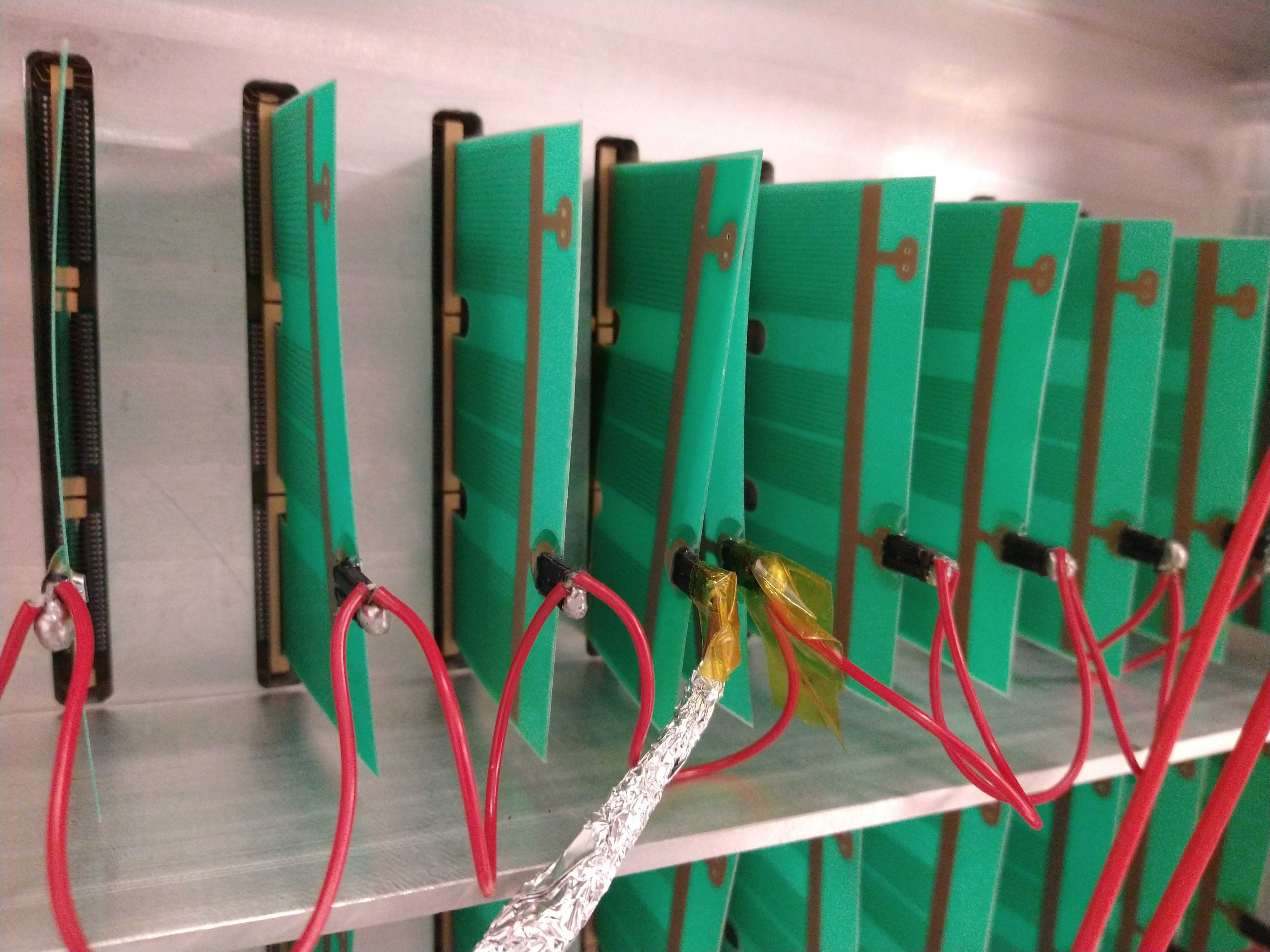}
\caption{\label{sec:setup:fig:shortingcard}Shorting cards are used to group pads from three connectors at the back of the OROC into a single channel. The fourth card from the left is passivated on the lower two legs and connects with the upper leg only to the top connector. These 21 pads are opposite to the \fe{55} source and read out as the signal channel. The following card connects to the middle and lower connector in this row and the corresponding pad groups are connected to ground, as are all other pad groups which are not read out.}
\end{figure}
The measurements taken for this paper investigate the response of the test OROC to an \fe{55} source. The source is suspended in the vessel close to the OROC wires, so that only a small part of the OROC can see the source; for this reason, we read out only a small section of the OROC pad plane. 
The pad plane is segmented into 468 groups of 21 or 22 pads, as indicated in \figref{sec:intro:subsec:ALICE:fig:OROC_Diagram}. To group pads for readout or for shortening to ground, we use \textit{shorting cards} (\textit{cf}. \figrefbra{sec:setup:fig:shortingcard}). These cards connect to the sockets on the back of the OROC and each connect three pad groups (21, 22 and 21 pads each) into one channel. Using one third of the connections of a shorting card, one group of 21 inner ($6\times\SI{10}{\milli\meter\squared}$) pads near the \fe{55} source are grouped and read out together as a single channel. We refer to this channel as the signal channel. To measure only background signals (cosmic radiation) and noise, another card is used in the same way to readout another pad group (21 inner pads) from a part of the pad plane away from the source. This channel is referred to as the background channel. The positions of these two pad groups are shown in \figref{sec:intro:subsec:ALICE:fig:OROC_Diagram}, all other pads are connected to ground.\\ 
{\indent}The signals from each channel are amplified by CREMAT CR-111 charge sensitive pre-amplifiers mounted in CREMAT CR-150-R5 evaluation boards. The pre-amplifiers have gains of $\SI{0.121}{\milli\volt\per\pico\coulomb}$ (signal channel) and $\SI{0.128}{\milli\volt\per\pico\coulomb}$ (background channel), and nominal output waveforms with a \SI{3}{\nano\second} rise-time and \SI{50}{\micro\second} decay. 
Full specifications of the pre-amplifiers are given in Ref. \cite{CR112SpecSheet}.\\
{\indent}When charge is induced on the OROC pad plane, the output from the preamplifier is a pulse with a height proportional to the induced charge. Following pre-amplification, a low pass filter is used to reduce noise and the output waveforms are then digitised by a CAEN N6730 digitiser with \SI{2}{\volt} dynamic range and a \SI{500}{\mega\hertz} sampling frequency. The shape of typical signals recorded by the digitiser are described in section \secref{sec:analysis} and an example of a typical waveform is shown in \figref{sec:analysis:genericWaveform}.

\subsection{Measurement procedure}

For each gas mixture and pressure, several voltage settings with varying anode voltage ($V_{\text{a}}$) but the same gating grid voltage ($V_{\text{gg}}$) and drift cathode voltage ($V_{\text{d}}$) are used. As we measured with an \fe{55} source, space charge distortions are of no concern and all gating grid wires were supplied with the same $V_{\text{gg}}$ without any additional potential, operating the chamber with the gating grid constantly open. The voltage at the gating grid wires is optimised to match the potential given by the drift field \SI{3}{\milli\meter} above the OROC's cathode wires ($\frac{\SI{-476}{\volt\per\centi\meter}}{\SI{3}{\milli\meter}} = \SI{-143}{V}$). 
In the test setup in the pressure vessel the edge wires are connected to ground, the rest of the anode wire grid is biased with a HV $V_{\text{a}}$. The anode grid is split into 7 HV regions, which are all biased by the same power supply (PS) channel. A protection resistor of \SI{3}{\mega\ohm} is installed between each region at the point where the supply lines are soldered together and from there fed to the PS. \\
\indent At each setting, several thousand waveforms are acquired, where the data acquisition (DAQ) triggers either on the channel integrating the signals from 21 pads at the location of the \fe{55} source (signal channel) or on the channel integrating the signals of the same number of pads in an area distant to these signal pads (background channel). \Figref{sec:analysis:genericWaveform} shows an example waveform for a pulse recorded by the DAQ. The trigger thresholds are optimised to be just below the pulse's baseline, allowing for noise events to be recorded, too. During data taking the digitiser is prevented from saturating and the trigger threshold is adjusted in cases where \textit{e.g.} a baseline change leads to a saturation of the DAQ with noise signals. If the readout is triggered by either the signal channel or background channel, both channels are read out. The recorded waveforms include a pre-trigger region in which one quarter of the waveform's samples are data recorded before the sample where the trigger condition is met (\textit{pre trigger}, \figrefbra{sec:analysis:genericWaveform:full}). Typically, waveforms are recorded for \SI{300}{\micro\second} including a pre-trigger time of \SI{75}{\micro\second}.\\
{\indent}After it is deemed that enough signals have been recorded for one setting (typically about 10,000 recorded waveforms), $V_{\text{a}}$ is changed and data is taken at the new configuration.

\paragraph{\arco{} data taking}$\ $\\
{\noindent}For the measurement with \arco{}, premixed \arcois{90-10}, grade N4, was used. This gas mix was used previously for the commissioning of the OROC for the ALICE TPC in 2009 \cite{Knichel}. Tests were first performed at atmospheric pressure inside the test box with an open gas system (this test box is described fully in Ref. \cite{Knichel}). The tests inside the high pressure vessel at RHUL are described in this paper. Pressures measured start at \SI{1}{barA} and are increased in steps of \SI{0.25}{barA} up to \SI{2.5}{barA}, and then in steps of \SI{0.5}{bar} up to \SI{4}{barA} -- a total of 10 pressure settings. To investigate repeatability, the vessel was then evacuated to vacuum and filled again with the same gas mix and additional data was taken at the pressure settings \SI{1.75}{barA}, \SI{3}{barA}, \SI{4}{barA}, and \SI{4.5}{barA}. \Tabref{sec:results:tab:measuredSettings:co2} lists the full list of \arco{} mixtures and pressures used.
For the data taken with premixed \arcois{90-10} gas mixture, the pressure was monitored in bar gauge (barG) and converted to barA assuming an atmospheric pressure of \SI{1}{bar}. To account for local variations in atmospheric pressure, these pressures are given with an error of \SI{35}{mbar}.

\paragraph{\arch{} data taking}$\ $\\
{\noindent}For the measurement campaign with \arch{}, premixed \archis{90-10} ($\pm \, \SI{0.2}{\%}$) was used and diluted with pure \ar{} of grade N5.5. We aimed for a mixture with no more than \SI{5}{\%} \ch{} to avoid safety issues related to flammability. Starting from about \SI{1}{barA} pressure, the gas mixture in the high pressure vessel was topped up with \archis{90-10} and \ar{} to increase the gas pressure in \SI{0.5}{bar} to \SI{1}{bar} steps. Increasing the pressure reduces the gain at a fixed voltage $V_{\text{a}}$, so the \ch{} fraction was slightly decreased during filling to increase the amplification factor.
In \tabref{sec:results:tab:measuredSettings:ch4} the \arch{} mixtures and pressures are listed in chronological order of the data taking. 
Since the \arch{} mixtures are mixed with the HPTPC gas system, measurements in barG with an assumed atmospheric pressure are no longer sufficient to provide an accurate measurement of the pressure and mixing ratio. 
Hence, for the data taking with \arch{}, an additional pressure sensor was used to measure the ambient atmospheric pressure outside of the vessel and determine an accurate measurement of the gas pressure in \si{barA}.
We estimate the uncertainty to be less than \SI{5}{\milli barA} and we use this value throughout the paper for the \arch{} measurements.
It is worth noting that the \SI{3.49}{barA} mixture with \SI{2.7(2)}{\%} \ch{} could not be operated in a stable manner. Increasing the \ch{} content during the next mixture and pressure change allowed for a stable operation at higher pressure once again.\\[0.1cm]
{\indent}The minimum voltage setting at each pressure for the \arco{} mixture was chosen to be the lowest voltage at which the gas gain could be measured, i.e. the lowest voltage at which the \fe{55} photopeak is visible in the \textit{Amplitude} spectrum above the trigger threshold. This is limited by electronic noise introduced in the readout chain.
The maximum voltage setting at each pressure is decided by the onset of current draw on the anode wires (or discharges), since operating at voltages higher than this results in voltage drops on the anode and risks of damaging the wires. \Tabref{sec:results:tab:measuredSettings} details the minimum and maximum voltage settings used for each gas mixture and pressure setting. Note that for the pressure settings \SI{1.25}{barA} and \SI{1.5}{barA} in \arco{} we were still unfamiliar with the limits of the anode, so a lower maximum voltage of \SI{1900}{\volt} was chosen, even when no current was observed. For later measurements all voltages below \SI{3000}{\volt} were considered, however, we did not reach this limit with any gas mixture before observing currents on the anode wires.
With \arch{}, we observe a lower ratio of \fe{55} signals compared to signals due to (e.g.) cosmic radiation. To account for this, more time is spent at each voltage setting (hence more waveforms are recorded.) Typically we record between 12,000 - 36,000 waveforms at each voltage setting with \arch{}, compared to between 4000 - 12000 waveforms with \arco{}.

\section{Data analysis}
\label{sec:analysis}

\begin{figure}
\centering
\subfloat[]{
\begin{tikzpicture}
  \node[anchor=south west,inner sep=0] (image) at (-0.3,0) {
  \includegraphics[height=0.25\textheight, trim = 0 0 0 0, clip=true]{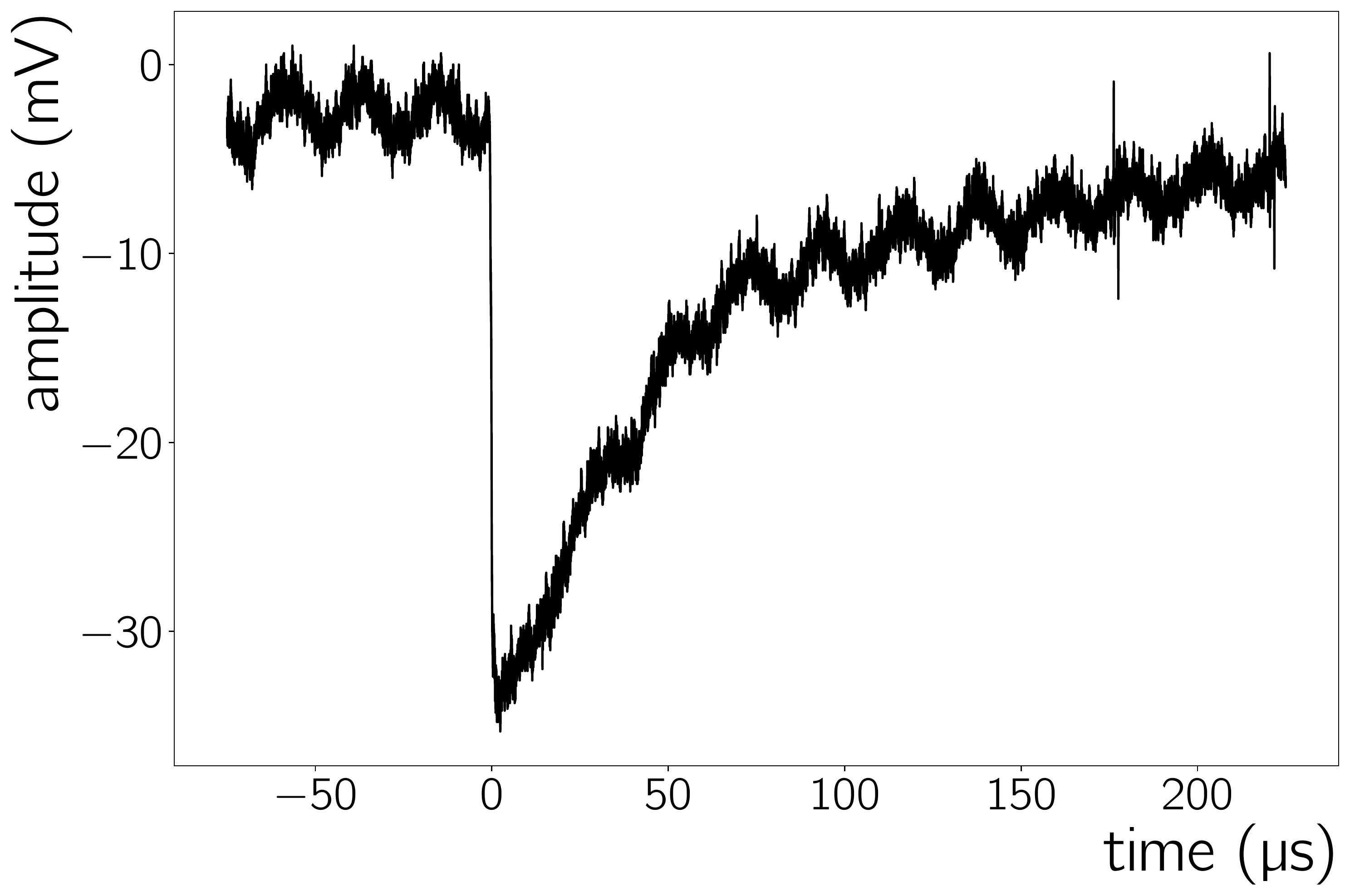}
  };
  \draw [ultra thick, blue, -] (1.30, 1.275) -- ++(7.7, 0) node [align=center, pos=0.5] {\raisebox{\baselineskip}{Amplitude}};
  \draw [ultra thick, blue, -] (1.30, 5.600) -- ++(7.7, 0) node [align=center, pos=0.5] {\raisebox{\baselineskip}{Baseline}};
  \draw [ultra thick, magenta, |-|]  (1.3, 3.200) -- ++(1.93, 0) node [align=center, pos=0.5] {};
  \node[magenta] (hello) at (2.25, 3.150) {\raisebox{-2.0\baselineskip}{Pre-trigger}};
  \node[magenta] (hullu) at (2.25, 2.750) {\raisebox{-2.0\baselineskip}{region}};
  \draw [thick, magenta, -]       (3.28, 0.500) -- ++(0, 6.5) node [align=center, pos=-0.05] {\rotatebox{90}{$t_{\text{peak}}$}};
\end{tikzpicture}
\label{sec:analysis:genericWaveform:full}
}   
\subfloat[]{
\begin{tikzpicture}
  \node[anchor=south west,inner sep=0] (image) at (-0.3,0) {
  \includegraphics[height=0.25\textheight, trim = 0 0 0 0, clip=true]{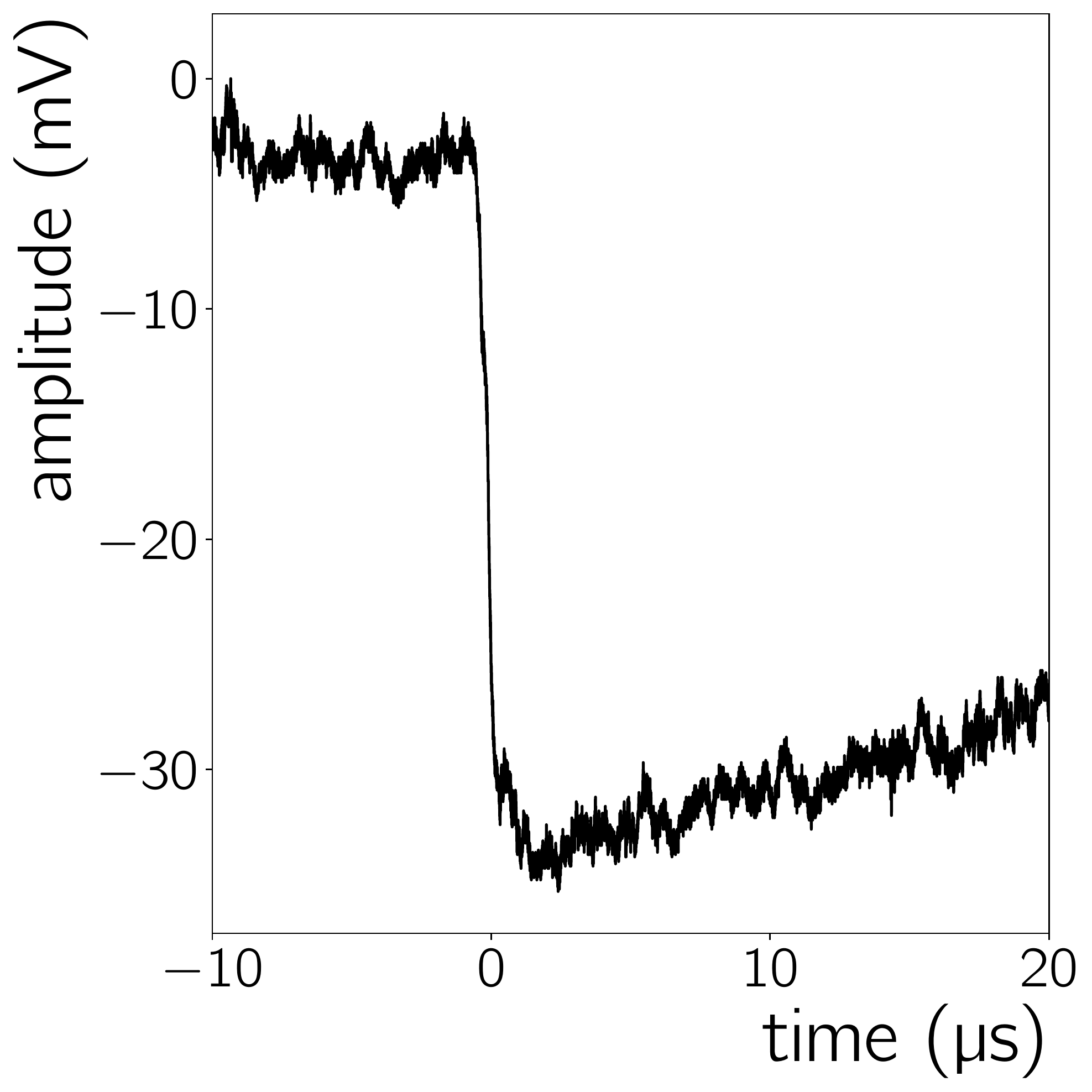}
  };
  \draw [ultra thick, blue, -]  (1.00, 1.275) -- ++(4.85, 0) node [align=center, pos=0.5] {\raisebox{-1.5\baselineskip}{Amplitude}};
  \draw [ultra thick, blue, -]  (1.00, 5.600) -- ++(4.85, 0) node [align=center, pos=0.5] {\raisebox{\baselineskip}{Baseline}};
  \draw [thick, magenta, -]       (3.0, 0.500) -- ++(0, 6.5) node [align=center, pos=-0.05] {\rotatebox{90}{$t_{\text{peak}}$}};
\end{tikzpicture}
\label{sec:analysis:genericWaveform:zoom}}
\caption{\label{sec:analysis:genericWaveform}A typical waveform recorded for the studies in this paper: \protect\subref{sec:analysis:genericWaveform:full} The full waveform and \protect\subref{sec:analysis:genericWaveform:zoom} a zoom into the region of the peak. Lines indicate the values the analysis identified for the baseline (\textit{Baseline}), the negative pulse amplitude (\textit{Amplitude}), and the time sample where the sample with the highest amplitude was found ($t_{\text{peak}}$). The oscillations on top of the waveform are electronic noise.}
\end{figure}
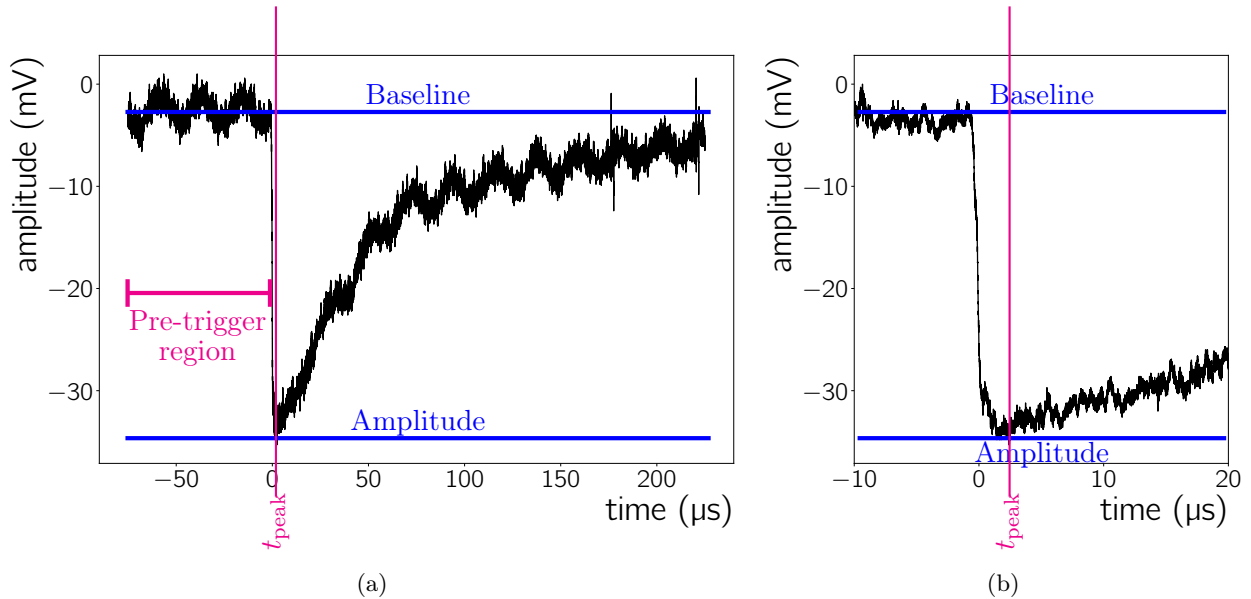
A typical example of a pulse recorded by the DAQ is shown in \figref{sec:analysis:genericWaveform}. For each waveform our analysis framework (described in Ref. \cite{instruments5020022}) calculates several characteristics. The most important ones for the analysis presented here are the \textit{Baseline} value, the \textit{Amplitude} value and the ``peak-time'', $t_{\text{peak}}$. The \textit{Baseline} is the mean of all samples in the pre-trigger window, \textit{i.e.} all samples for $t<0$. The \textit{Amplitude} corresponds to the pulse's negative amplitude: the sample with the lowest amplitude ($A_{\text{min}}$) is identified and the \textit{Amplitude} value is calculated as $\left|A_{\text{min}}-\textit{Baseline}\right|$. Furthermore, the time at which $A_{\text{min}}$ occurs is identified is $t_{\text{peak}}$.

\subsection{Analysis of Amplitude spectra}

\begin{figure}
\centering
\subfloat[$V_{\text{a}}=\SI{2150}{\volt}$]{\label{sec:analysis:fig:AmplitudeSpectra2150V}
\includegraphics[width=0.32\columnwidth, trim = 0 0 0 30, clip=true]{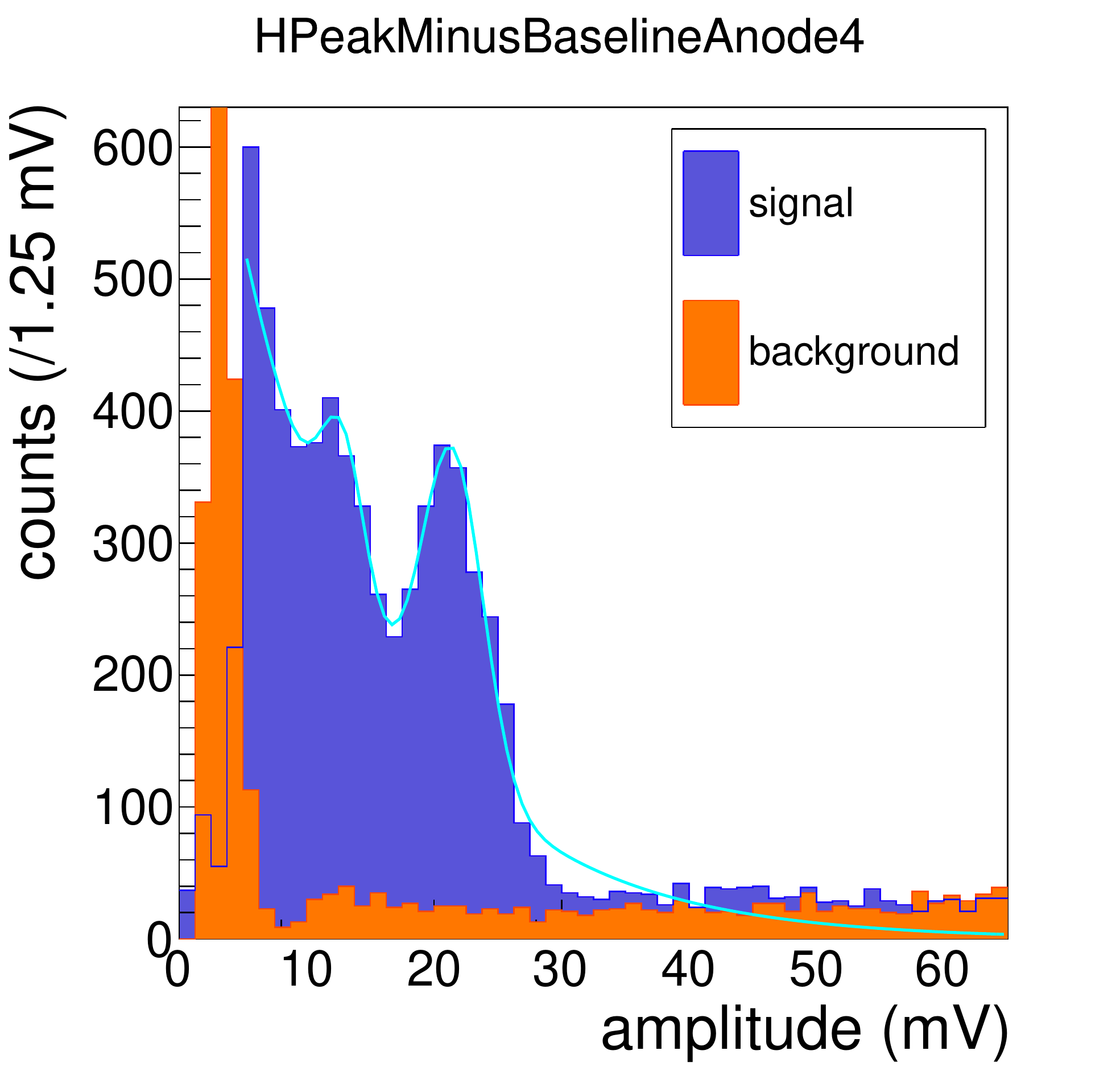}}
\subfloat[$V_{\text{a}}=\SI{2175}{\volt}$]{\label{sec:analysis:fig:AmplitudeSpectra2175V}
\includegraphics[width=0.32\columnwidth, trim = 0 0 0 30, clip=true]{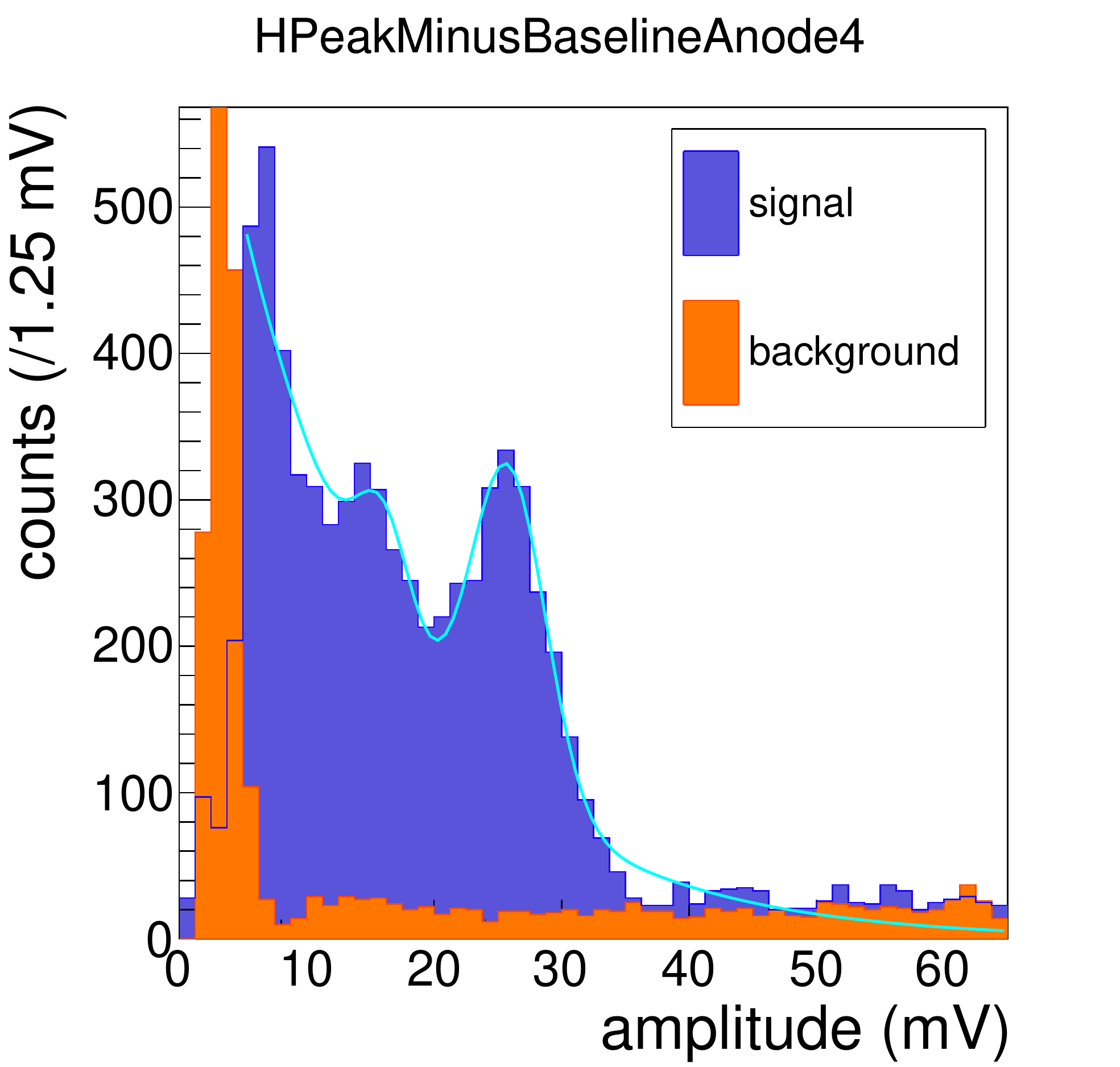}}
\subfloat[$V_{\text{a}}=\SI{2200}{\volt}$]{\label{sec:analysis:fig:AmplitudeSpectra2200V}
\includegraphics[width=0.32\columnwidth, trim = 0 0 0 30, clip=true]{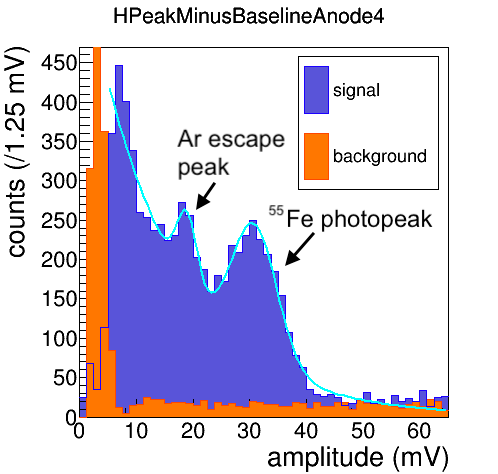}}
\caption{\label{sec:analysis:fig:AmplitudeSpectra}\textit{Amplitude} spectra for data taken at \SI{1750}{\milli bar} absolute pressure \arcois{90-10} with $V_{\text{gg}}=-\SI{143}{\volt}$ and $V_{\text{d}}=-\SI{16}{\kilo\volt}$ ($E_{\text{d}}=\SI{-476}{\volt\per\centi\meter}$). The spectra are for data taken with anode voltages of \protect\subref{sec:analysis:fig:AmplitudeSpectra2150V} \SI{2150}{\volt}, \protect\subref{sec:analysis:fig:AmplitudeSpectra2175V} \SI{2175}{\volt} and \protect\subref{sec:analysis:fig:AmplitudeSpectra2200V} \SI{2200}{\volt} after cleaning cuts have been applied, blue and orange histograms show signal and background channels respectively. The photopeak of the \fe{55} source's x-ray emission can be seen in the each spectra at \protect\subref{sec:analysis:fig:AmplitudeSpectra2150V} \SI{21}{\milli\volt}, \protect\subref{sec:analysis:fig:AmplitudeSpectra2175V} \SI{26}{\milli\volt} and \protect\subref{sec:analysis:fig:AmplitudeSpectra2200V} \SI{30}{\milli\volt} and the lower amplitude $\text{Ar}$ escape-peak at \SI{11}{\milli\volt}, \SI{14}{\milli\volt} and \SI{18}{\milli\volt}. A fit to this spectrum according to Eq.~\eqref{sec:analysis:eq:arco2fit} is shown in each plot.}
\end{figure}
\begin{figure}
\centering
\subfloat[\arcois{90-10}]{\label{sec:analysis:fig:arco2fits:peakPos}\includegraphics[height=0.25\textheight,trim = 0 0 0 45, clip=true]{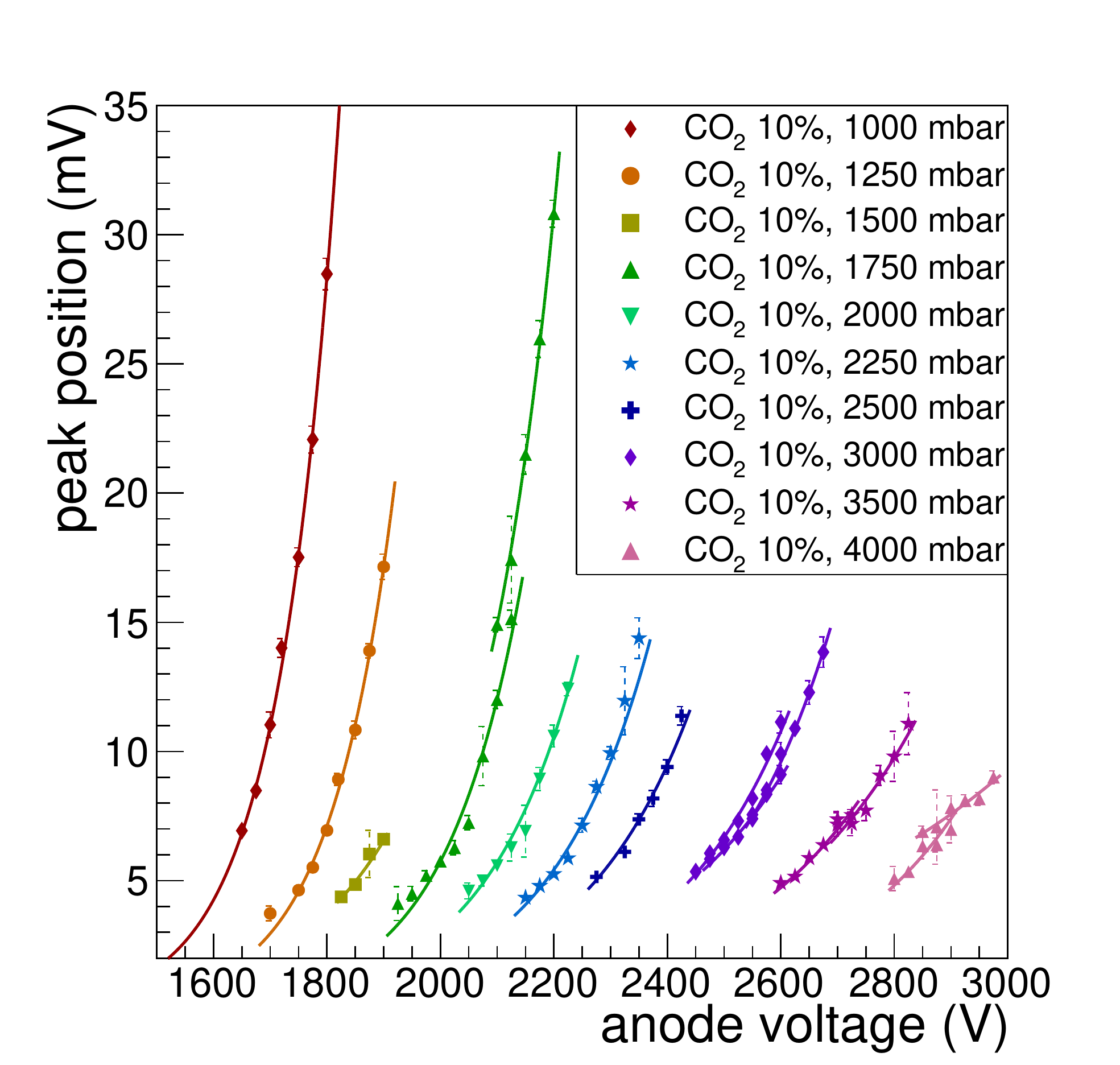}}
\subfloat[\arch{}]{\label{sec:analysis:fig:arch4fits:peakPos}\includegraphics[height=0.25\textheight]{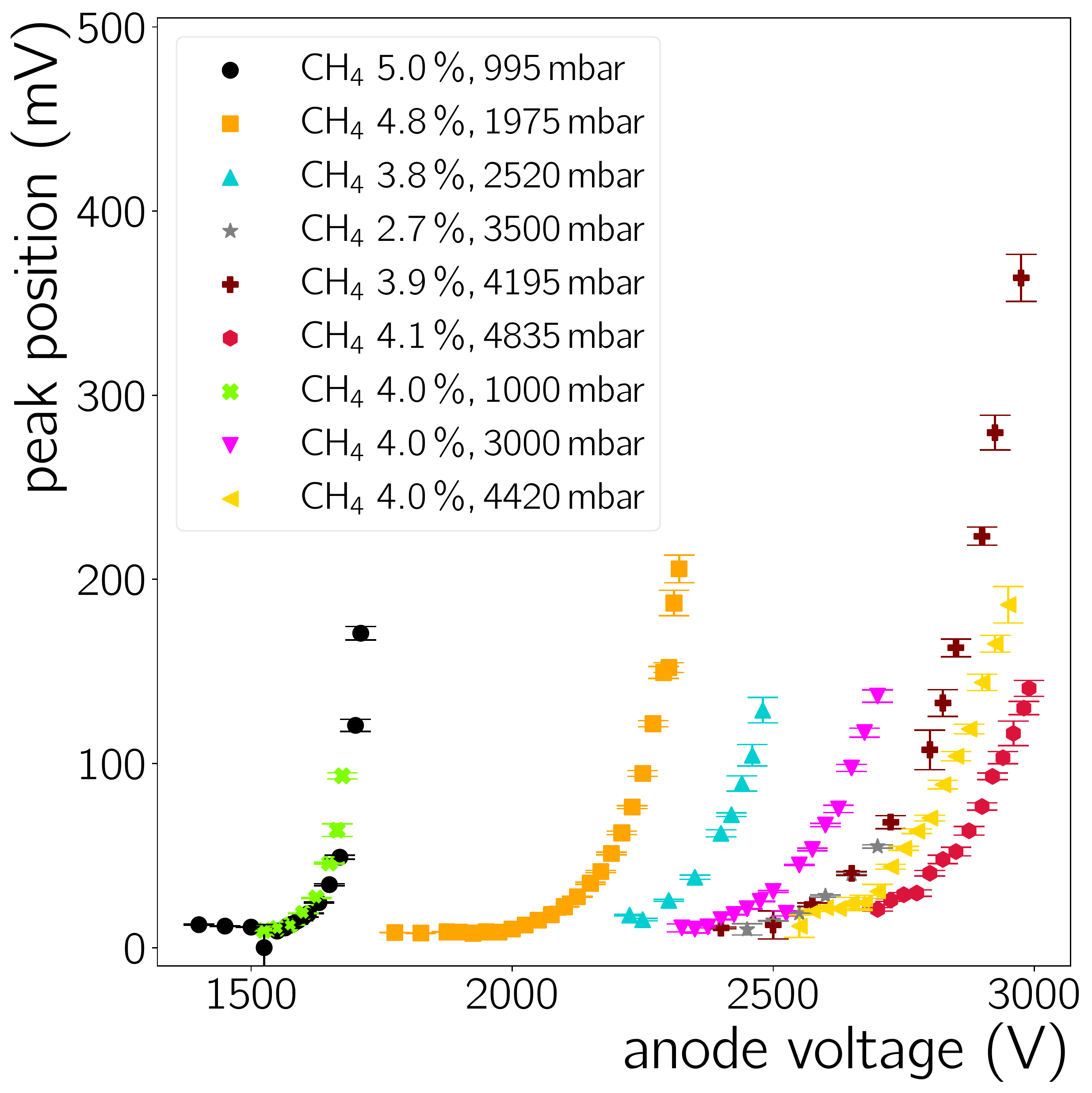}}
\\
\subfloat[\arcois{90-10}]{\label{sec:analysis:fig:arco2fits:peakSigma}\includegraphics[height=0.25\textheight,trim = 0 0 0 45, clip=true]{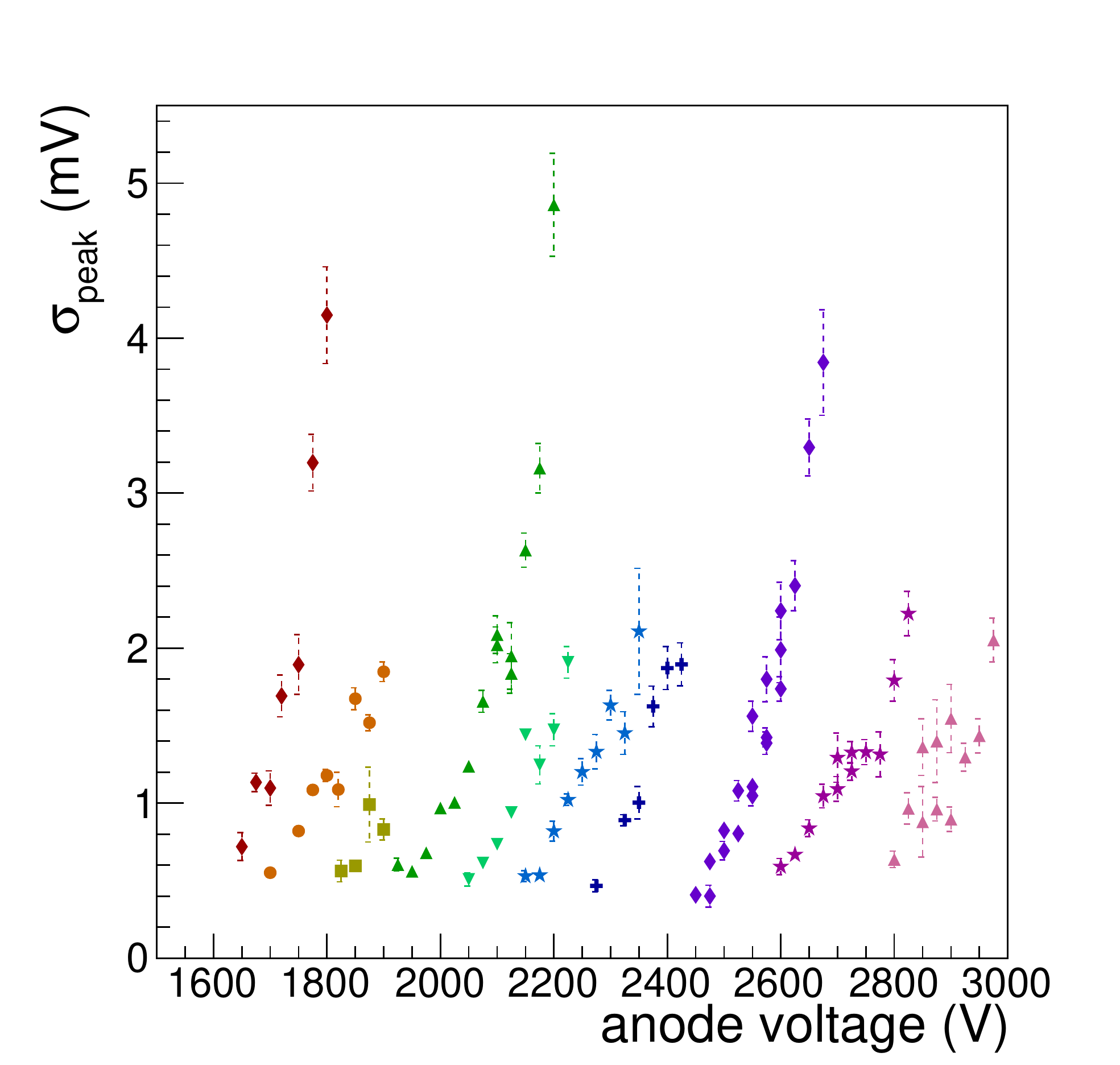}}
\subfloat[\arch{}]{\label{sec:analysis:fig:arch4fits:peakSigma}\includegraphics[height=0.25\textheight]{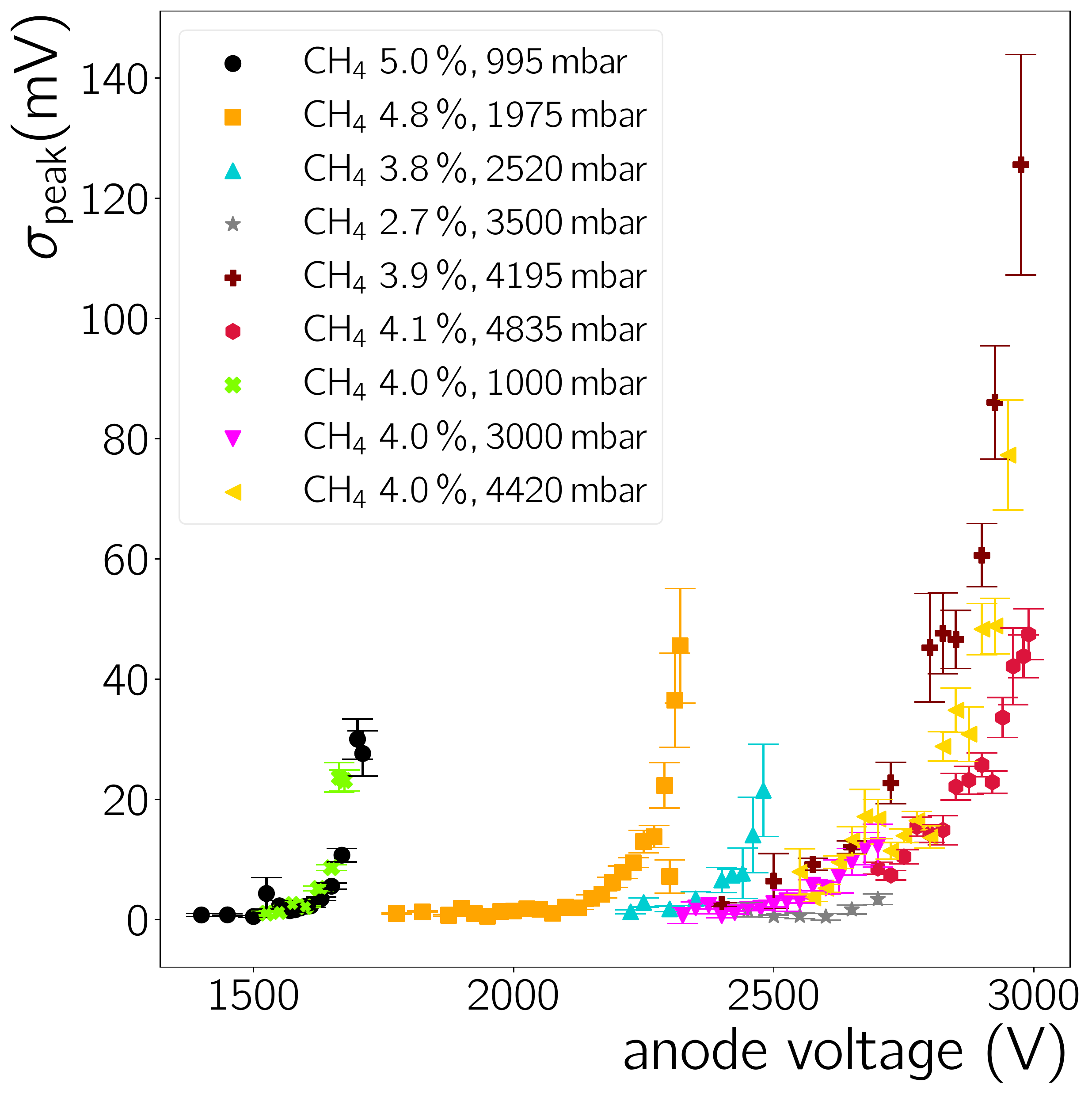}}
\caption{\label{sec:analysis:fig:aco2andch4fits}Fit results for fits of Eq.~\eqref{sec:analysis:eq:arco2fit} and Eq.~\eqref{sec:analysis:eq:arch4fit} to the \arcois{90-10} (first column) and \arch{} (second column) data, respectively. In the first row, \protect\subref{sec:analysis:fig:arco2fits:peakPos} and \protect\subref{sec:analysis:fig:arch4fits:peakPos}, the photopeak position ($p_6$ and $p_3$ in Eq.~\eqref{sec:analysis:eq:arco2fit} and Eq.~\eqref{sec:analysis:eq:arch4fit}, respectively) is shown. The plots in the second row, \protect\subref{sec:analysis:fig:arco2fits:peakSigma} and \protect\subref{sec:analysis:fig:arch4fits:peakSigma}, display the fit result for  $\sigma_{\text{peak}}$ ($p_7$ and $p_4$ in \eqref{sec:analysis:eq:arco2fit} and \eqref{sec:analysis:eq:arch4fit}, respectively). The data was taken with the OROC using $V_{\text{gg}}=-\SI{143}{\volt}$ and $V_{\text{d}}=-\SI{16}{\kilo\volt}$ ($E_{\text{d}}=\SI{-476}{\volt\per\centi\meter}$), whilst varying the anode voltage. See \tabref{sec:results:tab:measuredSettings} for the uncertainties on pressure values and mixing ratios.}
\end{figure}

Two cuts are applied on the calculated waveform parameters. The first cut requires the \textit{Baseline} to be between $-\SI{10}{\milli\volt}$ and $\SI{10}{\milli\volt}$. This removes events triggered by baseline fluctuations caused by \textit{e.g.} discharges. The second cut requires that $\SI{0.5}{\micro\second} <  t_{\text{peak}}$. This cut on $t_{\text{peak}}$ ensures that only pulses with a sensible shape are considered. We observe incidental noise peaks with a width of a few \SI{100}{\nano\second}, which are removed by this cut. 
Exponential smoothing is applied to the waveforms to counteract small fluctuations (RMS of $\le \SI{3}{\milli\volt}$.) Each waveform is corrected according to:
\begin{align*}
    V_i^{\textrm{smooth}} = \alpha_{\textrm{smooth}} \cdot V_i + (1 - \alpha_{\textrm{smooth}}) \cdot V_{i-1}.
\end{align*}
Where $V_i^{\textrm{smooth}}$ and $V_i$ are the $i^{\text{th}}$ sample (at time $t_i$) of the smoothed and raw waveform, respectively. The factor $\alpha_{\textrm{smooth}}$ is chosen as 0.005. A full discussion of the waveform smoothing as well as examples of smoothed waveforms can be found in Ref. \cite{instruments5020022}. Further analysis is then performed on histograms of \textit{Amplitude} values for signals recorded at each voltage, gas mixture and pressure setting.\\
{\indent}\Figref{sec:analysis:fig:AmplitudeSpectra} shows three \textit{Amplitude} spectra for different $V_{\text{a}}$ measured in \arcois{90-10} at \SI{1.75}{barA}. 
Comparing the signal and background channel spectra in these plots allows us to identify the background component due to cosmic radiation and the signal component from the \fe{55} x-rays. 
For spectra such as those shown in \Figref{sec:analysis:fig:AmplitudeSpectra}, the photopeak of the \fe{55} source's x-ray emission and the $\text{Ar}$ escape-peak are both visible. To extract parameters related to both of these peaks, the \textit{Amplitude} spectra are fitted with the function
\begin{align}
f_1\left({Amplitude}\right) = \text{e}^{p_0 + p_1\cdot {Amplitude}} + p_2 \cdot \text{e}^{-0.5 \cdot {\left(\frac{{Amplitude} - p_3}{p_4}\right)}^2} + p_5 \cdot \text{e}^{-0.5 \cdot {\left(\frac{{Amplitude} - p_6}{p_7}\right)}^2} \quad.
\label{sec:analysis:eq:arco2fit}
\end{align}
The first term of $f_1$ is an exponential function to fit the background and the second and third terms are Gaussian functions to fit the photopeak and escape-peak. \Figref{sec:analysis:fig:AmplitudeSpectra} includes this function fitted to the signal channel's \textit{Amplitude} spectra.\\
{\indent}The escape-peak cannot be seen in all spectra. At lower OROC gas gains, the escape-peak has low enough amplitude such that it overlaps with the background exponential, and for higher observed OROC gas gains, the escape-peak is often washed out by the broadening of the peaks in the spectra. This is particularly true in the case of the \arch{} data, where we observe a lower ratio of \fe{55} signals as compared to background signals. This change in ratio may be due to the about $\SI{6}{\%}$ lower $W$ value in \arch{} mixtures (\ch{} contents of $\sim\!\!\SI{4}{\%}$) as compared to \arcois{90-10} \cite{kolanoski2016teilchendetektoren}, while the number of electron-ion pairs produced by \textit{e.g.} cosmic radiation is comparable ($\sim\!\!\SI{35}{pairs\per\centi\meter}$) for both gas mixtures \cite{garfieldpp,Biagi1018382,SMIRNOV2005474}. In addition, the transversal diffusion in \archis{96-4} as compared to \arcois{90-10} is more than 3 times larger in the former \cite{Biagi1018382}. Thus the fraction of \fe{55} events, which are not fully contained in the instrumented pads, will be larger in the \arch{} mixtures studied and hence the \fe{55} peaks in the spectrum broader and of lower amplitude. For these reasons we also observe a worse energy resolution as compared to the \arco{} measurements, as will be shown in the next section. The challenges in fitting Eq.~\eqref{sec:analysis:eq:arco2fit} to the \arch{} data can be addressed by tuning the start parameters and fit range. As we are however primarily interested in the position of the \fe{55} photopeak, a simplified approach was used for this analysis: the spectra are instead fitted with the function:
\begin{align}
f_2\left({Amplitude}\right) = \text{e}^{p_0 + p_1\cdot {Amplitude}} + p_2 \cdot \text{e}^{-0.5 \cdot {\left(\frac{{Amplitude} - p_3}{p_4}\right)}^2} \quad,
\label{sec:analysis:eq:arch4fit}
\end{align}
where only the \fe{55} photopeak is fitted. Using this simplified approach of fitting $f_2$, \textit{i.e.} Eq.~\eqref{sec:analysis:eq:arch4fit}, allows for automated fitting of all measured voltages, \arch{} mixtures and pressure settings.\\
{\indent}\Figref{sec:analysis:fig:aco2andch4fits} shows the fit results for two of the free parameters of $f_1$ and $f_2$ for the fits to all \arco{} and \arch{} data. The $\chi^2$ divided by the number of degrees of freedom was between 0.8 and 5.4 for all fits, and typically between 0.8 and 1.8.

\section{Measurements and results}
\label{sec:results}

The peak position vs $V_{\text{a}}$ measurements (\figrefbra{sec:analysis:fig:arco2fits:peakPos}, \figrefbra{sec:analysis:fig:arch4fits:peakPos}) demonstrate already one main aim of this work: An ALICE OROC can be operated at up to at least five times its design pressure. The exponential trends observed with increasing $V_{\text{a}}$ for each gas mixture setting match the expectations given by the avalanche nature of the charge amplification. For a full characterisation of the chamber at high pressure, the gas gain and energy resolution are calculated from the fits to the \textit{Amplitude} spectra discussed in \secref{sec:analysis}.

\subsection{Energy resolution}
\label{sec:results:subsec:energyresolution}

\begin{figure}
\centering
\subfloat[\arcois{90-10}]{\label{sec:results:fig:EnergyRes_AllPressures:arco2}\includegraphics[height=0.25\textheight,trim=0 0 0 45,clip=true]{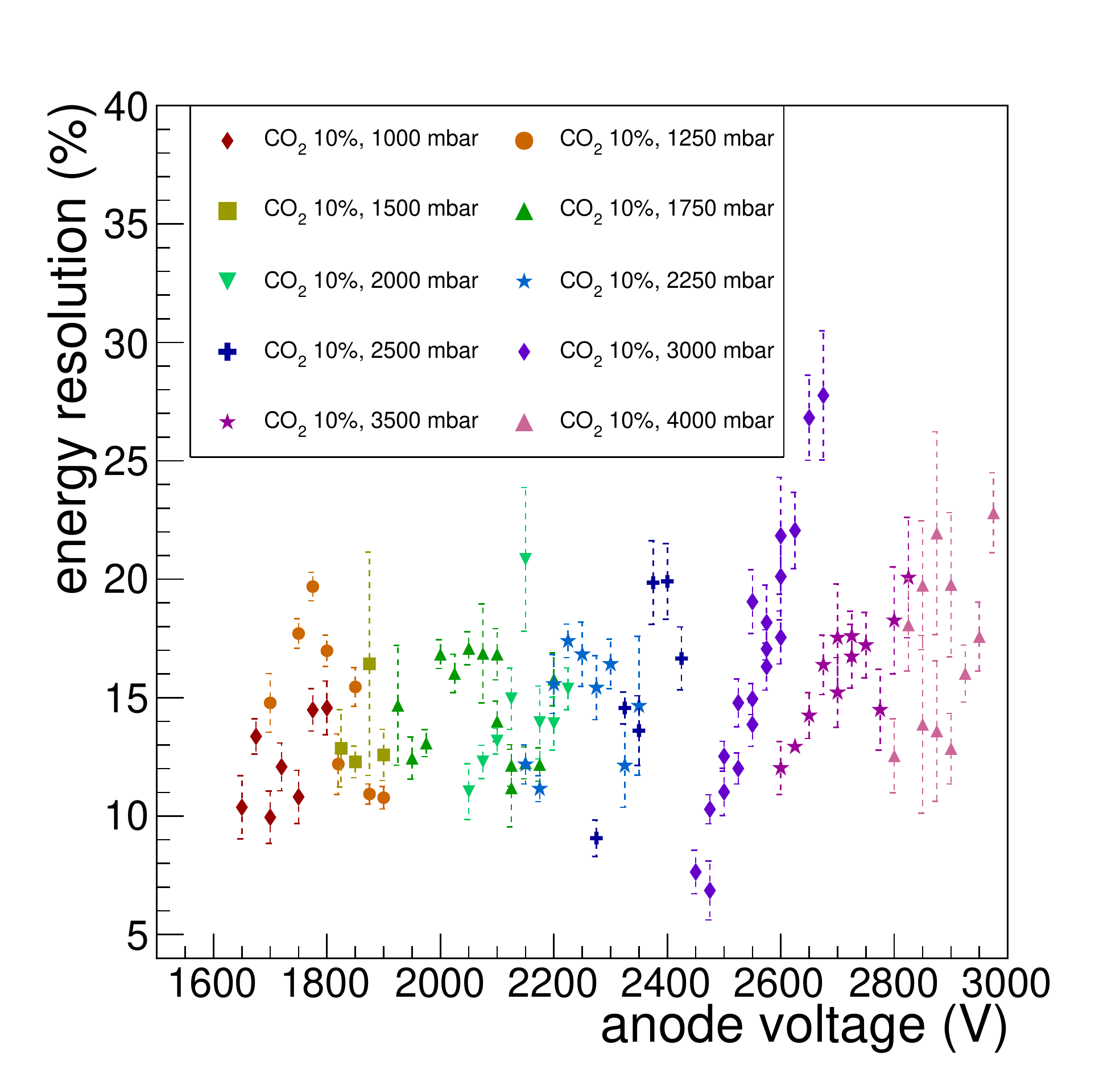}}
\subfloat[\arch{}]{\label{sec:results:fig:EnergyRes_AllPressures:arch4}\includegraphics[height=0.25\textheight]{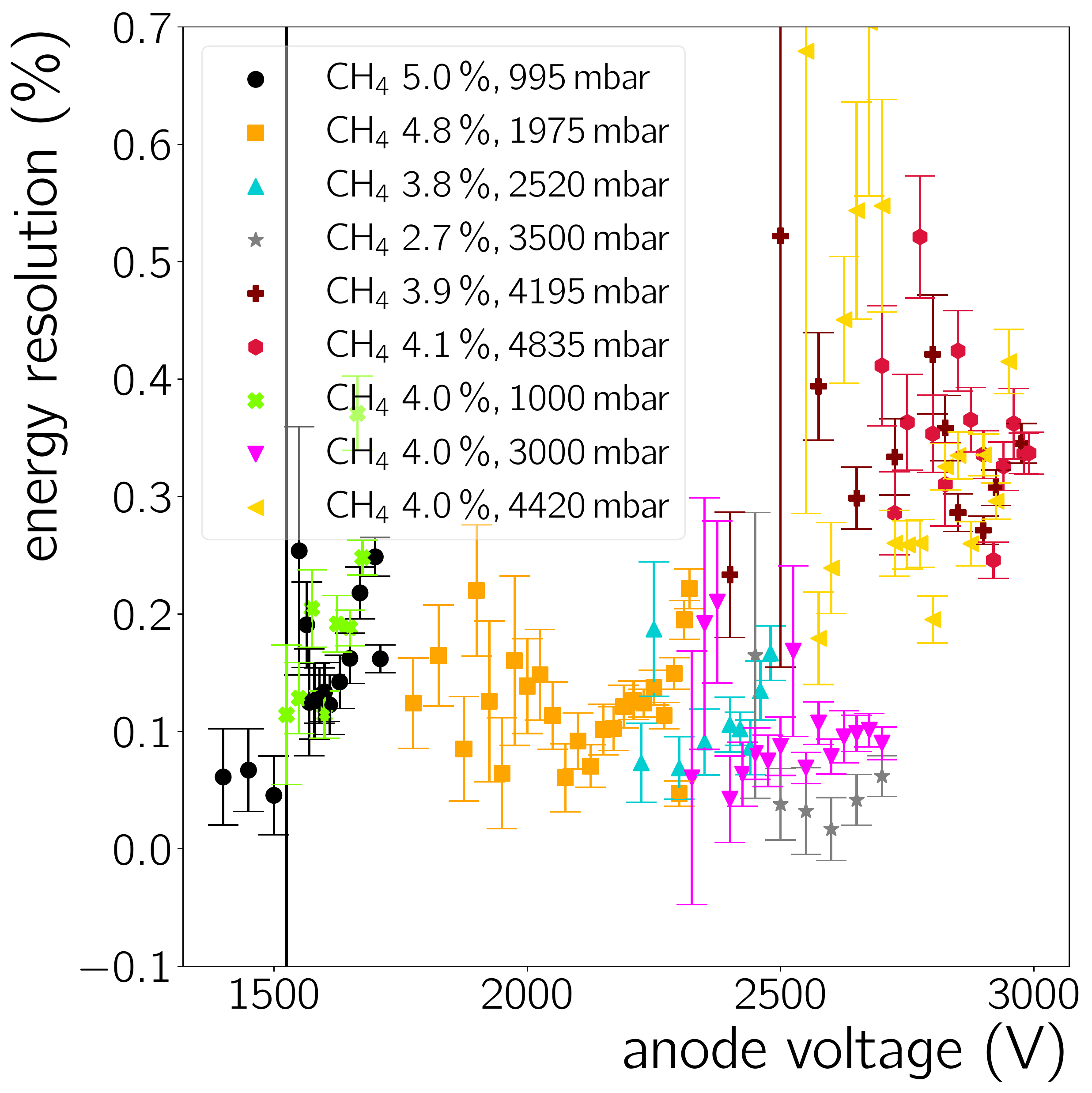}}
\caption{\label{sec:results:fig:EnergyRes_AllPressures}Energy resolution ($\frac{\sigma_{\text{peak}}}{\text{peak}\ \text{position}}$) as function of anode voltage for data obtained with an \fe{55} source for \protect\subref{sec:results:fig:EnergyRes_AllPressures:arco2} pressures between \SI{1}{barA} and \SI{4}{barA} in \arcois{90-10}, and \protect\subref{sec:results:fig:EnergyRes_AllPressures:arch4} pressures between \SI{1}{barA} and \SI{4.8}{barA} in different \arch{} mixtures. This data was taken with the OROC using $V_{\text{gg}}=-\SI{143}{\volt}$ and $V_{\text{d}}=-\SI{16}{\kilo\volt}$ ($E_{\text{d}}=\SI{-476}{\volt\per\centi\meter}$), whilst varying the anode voltage $V_{\text{a}}$.}
\end{figure}
\Figref{sec:results:fig:EnergyRes_AllPressures} shows the energy resolution ($\frac{\sigma_{\text{peak}}}{\text{peak}\ \text{position}}$) as a function of anode voltage. For the most part the value is between $\SI{10}{\%}$ and $\SI{20}{\%}$, $\SI{25}{\%}$ for the highest pressure measurements in \arch{} mixtures.
Better values than this have been obtained with OROCs, the performance observed in this paper is likely explained by two differences in our set-up:
The first is the alignment of the \fe{55} source, the readout integrates signals from a group of 21 pads which may not be perfectly aligned with the source.
Secondly, the pre-amplifiers are located outside of the high pressure vessel, this means that all signals have to travel through a cable of significant length before they reach the pre-amplifiers. \Figref{sec:analysis:genericWaveform} gives an idea of the noise encountered along the cable path.
Note that the measurements of energy resolution reported here refer specifically to the measurement with an \fe{55} source and are not a reflection of the ultimate energy resolution for tracked particles in the detector. 

\subsection{Gas gain}
\label{sec:results:subsec:gain}

\begin{figure}
\centering
\subfloat[\arcois{90-10}]{\label{sec:results:fig:GainVsVoltage_AllPressures:arco2}\includegraphics[height=0.25\textheight,trim = 0 0 0 50, clip=true]{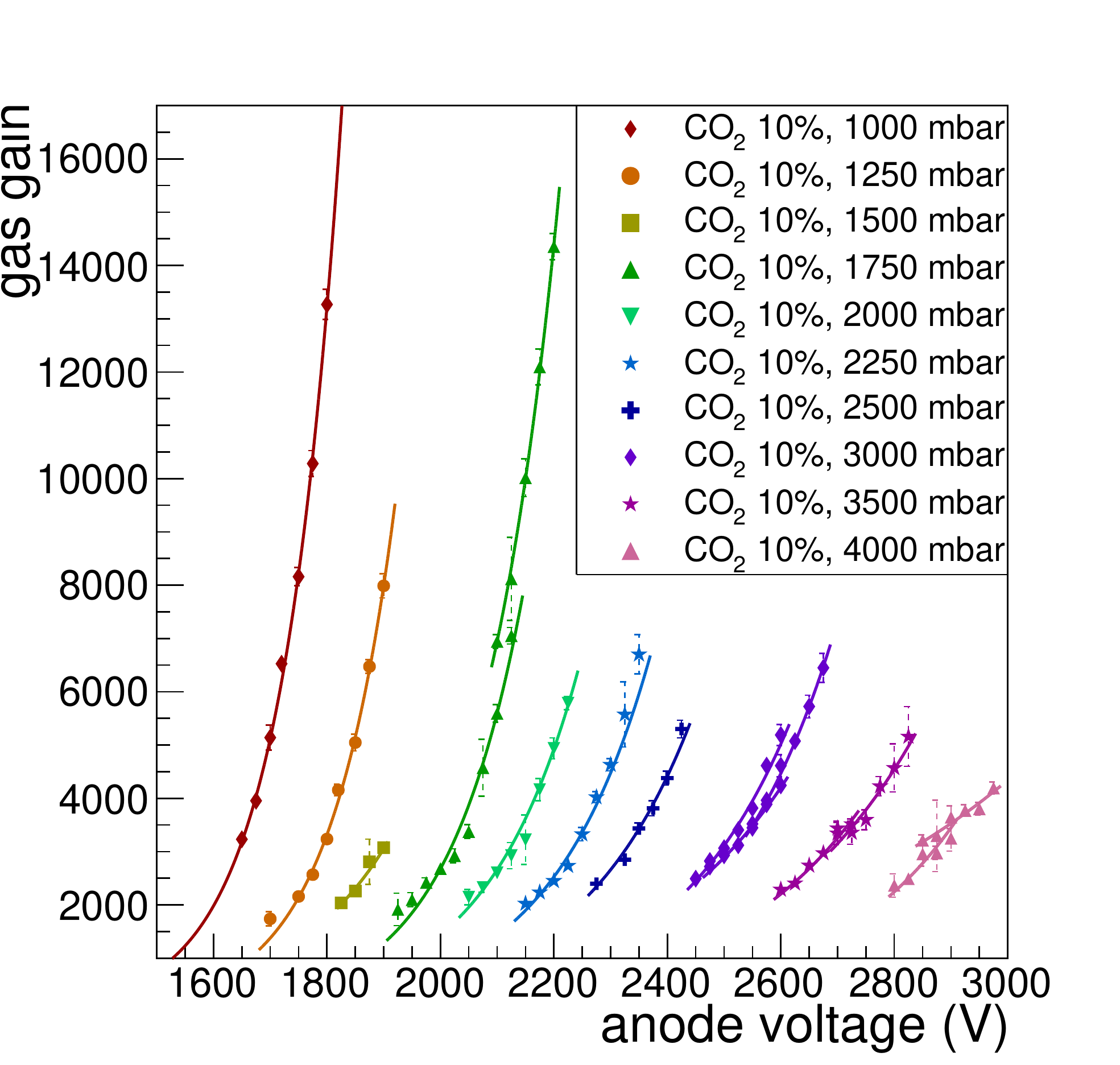}}
\subfloat[\arch{}]{\label{sec:results:fig:GainVsVoltage_AllPressures:arch4}\includegraphics[height=0.25\textheight]{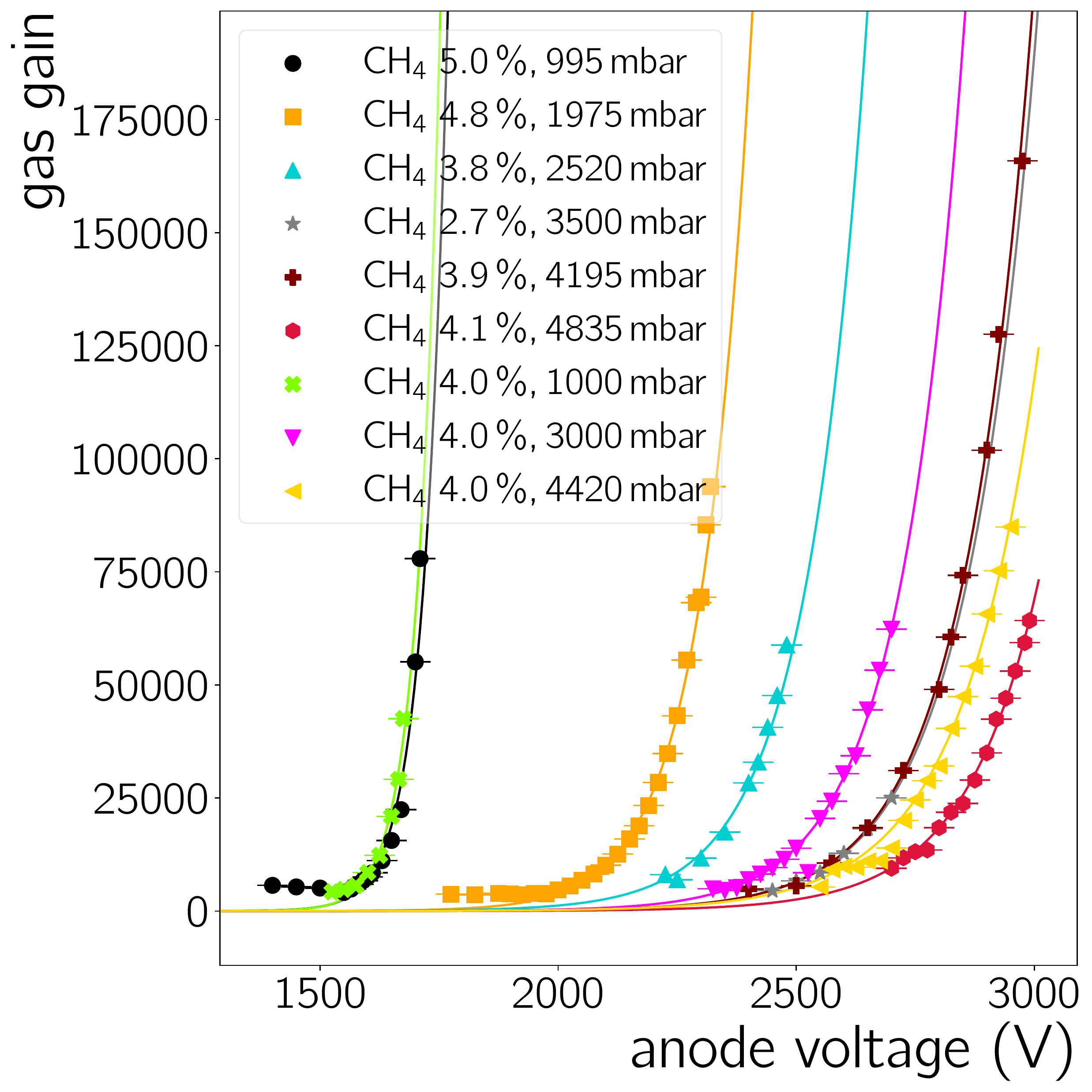}}
\caption{\label{sec:results:fig:GainVsVoltage_AllPressures}Gas gain vs anode voltage for \protect\subref{sec:results:fig:GainVsVoltage_AllPressures:arco2} pressures between \SI{1}{barA} and \SI{4}{barA} in \arcois{90-10}, and \protect\subref{sec:results:fig:GainVsVoltage_AllPressures:arch4} pressures between \SI{1}{barA} and \SI{4.8}{barA} in different \arch{} mixtures. All data has been obtained with an \fe{55} source. This data was taken with the OROC using $V_{\text{gg}}=-\SI{143}{\volt}$ and $V_{\text{d}}=-\SI{16}{\kilo\volt}$ ($E_{\text{d}}=\SI{-476}{\volt\per\centi\meter}$), whilst varying the anode voltage $V_{\text{a}}$. See \tabref{sec:results:tab:measuredSettings} for the uncertainties on pressure values and mixing ratios.}
\end{figure}

The gas amplification factor, or gas gain, $G_\text{OROC}$, can be calculated as
\begin{equation}
G_\text{OROC} = \frac{q_{\text{amp}}}{N_\text{e} \times q_{\text{e}}},\quad\text{where}\quad
q_{\text{amp}} = \frac{A}{G_{\text{preamp}}}\quad\text{and}\quad 
N_\text{e} = \frac{\varepsilon_{\text{particle}}}{W_{ij}}.    
\label{sec:results:subsec:gain:eq:gain}
\end{equation}
Here, $\varepsilon_{\text{particle}}$ is the energy deposited by some radiation inside the gas and $W_{ij}$ is the average energy to produce an electron-ion pair in a mixture of the gases $i$ and $j$, which makes $N_\text{e}$ the number of electrons liberated by the incoming radiation. 
The charge after amplification in the gas, \textit{i.e.} the input charge at the preamplifier is $q_\text{amp}$, $A$ is the amplitude of a measured signal, and $G_{\text{preamp}}$ denotes the preamplifier gain. The factor $q_{\text{e}}$ is the electron charge. 
For the CREMAT amplifiers used in this work, $G_{\text{preamp}}=\SI{0.121}{\milli\volt\per\pico\coulomb}$. $W_{ij}$ needs to be calculated for each gas mixture, according to
\begin{equation}
W_{ij} = \left(\frac{f_i}{W_i} + \frac{f_j}{W_j}\right)^{-1} \quad .
\label{sec:results:subsec:gain:eq:Wij}
\end{equation}
Where $f_i$ and $f_j$ are the mixing fractions of each component of the mixture. For example, $W_{\text{Ar-CO}_2}^{(90\text{-}10)}=\SI{26.6}{\electronvolt}$. The values for pure \ar{}, \co{} and \ch{} are $W_{\text{Ar}}=\SI{26}{\electronvolt}$, $W_{\text{CO}_2}=\SI{33}{\electronvolt}$, and $W_{\text{CH}_4}=\SI{28}{\electronvolt}$, respectively \cite{kolanoski2016teilchendetektoren}.\\
{\indent}The mean of the Gaussian function fitted to the photopeak, i.e. $p_6$ in Eq.~\eqref{sec:analysis:eq:arco2fit} and $p_3$ in Eq.~\eqref{sec:analysis:eq:arch4fit}, is taken as the mean energy deposit of the x-ray emission from the \fe{55} source, that is $A$ in Eq.~\eqref{sec:results:subsec:gain:eq:gain}. Hence, following the analysis in \secref{sec:analysis}, $G_\text{OROC}$ can be calculated for each anode voltage, pressure, and gas mixture. \Tabref{sec:results:tab:measuredSettings} shows the maximum stable voltages achieved in each gas mix and pressure, and the maximal $G_\text{OROC}$, denoted as $G^{\text{max}}$, measured at those settings. 
The OROC gas gain plotted as a function of voltage at each gas setting is shown in \figref{sec:results:fig:GainVsVoltage_AllPressures}.\\
\begin{table}
\centering
\subfloat[\arcois{90-10}]{\label{sec:results:tab:measuredSettings:co2}
\begin{tabular}[t]{c|c|c|c}
Pressure           & $V_{\text{a}}^{\text{min}}$ & $V_{\text{a}}^{\text{max}}$  & $G^{\text{max}}$ \\
(\si{\milli barA}) & (\si{\volt})                & (\si{\volt})                 & $\times 10^3$     \\ \hline
1000 & 1650 & 1800 & $13.3 \pm 0.3$ \\
1250 & 1700 & 1900 & $8.0 \pm 0.2$ \\
1500 & 1825 & 1900 & $3.1 \pm 0.1$ \\
1750 & 1925 & 2125 & $7.1 \pm 0.1$ \\
2000 & 2050 & 2225 & $5.8 \pm 0.1$ \\
2250 & 2150 & 2350 & $6.7 \pm 0.3$ \\
2500 & 2275 & 2425 & $5.3 \pm 0.1$ \\
3000 & 2475 & 2600 & $4.2 \pm 0.2$ \\
3500 & 2600 & 2725 & $3.5 \pm 0.1$ \\
4000 & 2800 & 2875 & $3.5 \pm 0.2$ \\ \hline
1750 & 2100 & 2200 & $14.4 \pm 0.3$ \\
3000 & 2450 & 2600 & $5.2 \pm 0.2$ \\
3000 & 2550 & 2675 & $6.5 \pm 0.3$ \\
3500 & 2700 & 2825 & $5.3 \pm 0.3$ \\
4000 & 2850 & 2975 & $4.2 \pm 0.1$ \\
\end{tabular}
}
\subfloat[\arch{}]{\label{sec:results:tab:measuredSettings:ch4}
\begin{tabular}[t]{c|c|c|c|cc}
Pressure           & $V_{\text{a}}^{\text{min}}$ & $V_{\text{a}}^{\text{max}}$  & $f_{\text{CH}_4}$ & $G^{\text{max}}$ & \\
(\si{\milli barA}) & (\si{\volt})                & (\si{\volt})                 & ($\si{\%}$)       & $\times 10^3$  & \\ \hline
 995 &  1400 & 1710 & $5.0\pm0.5$ &  $78\pm2$ & \\ 
1975 &  1775 & 2320 & $4.8\pm0.3$ &  $94\pm3$ & \\
2520 &  2225 & 2480 & $3.8\pm0.2$ &  $59\pm3$ & \\ 
3500 &  2450 & 2700 & $2.7\pm0.2$ &  $25\pm1$ & $^{\dagger}$ \\
4195 &  2400 & 2975 & $3.9\pm0.1$ & $166\pm6$ & \\ 
4835 &  2700 & 2990 & $4.1\pm0.1$ &  $64\pm2$ & \\ \hline
1000 &  1525 & 1675 & $4.0\pm0.5$ &  $43\pm1$ & \\
3000 &  2300 & 2675 & $4.0\pm0.2$ &  $62\pm2$ & \\
4420 &  2550 & 2975 & $4.0\pm0.1$ &  $85\pm5$ & \\
\end{tabular}}
\caption{\label{sec:results:tab:measuredSettings}\protect\subref{sec:results:tab:measuredSettings:co2}  All pressure settings for the \arcois{90-10} measurements, and pressure and \protect\subref{sec:results:tab:measuredSettings:ch4} gas mixture settings for the different \arch{} measurements with varying \ch{} fraction $f_{\text{CH}_4}$. The uncertainty on the pressure in \protect\subref{sec:results:tab:measuredSettings:co2} is $\pm\SI{30}{\milli\bar}$ and $\pm\SI{5}{\milli\bar}$ for the data in \protect\subref{sec:results:tab:measuredSettings:ch4}. The range of anode voltages measured for each setting from the smallest ($V_{\text{a}}^{\text{min}}$) to largest ($V_{\text{a}}^{\text{max}}$) for all settings is given, too. The setting marked by $\dagger$ in \protect\subref{sec:results:tab:measuredSettings:ch4} was not stable and no reasonable gain curve could be extracted. The horizontal line in each table splits two different periods of data taking with an evacuation of the vessel in-between.}
\end{table}
{\indent}At the same pressure and voltage settings, a much higher gas gain was measured with \arch{} than with \arco{}.
For example, in \arcois{90-10} at \SI{3}{barA} at $V_{\text{a}} = \SI{2675}{\volt}$, $G_\text{OROC}$ was measured to be $(6.5 \pm 0.3) \times 10^3$.
For the same pressure and anode voltage in \archis{96-4}, $G_\text{OROC}$ was measured to be $(62 \pm 2) \times 10^3$, a factor of 9.5 greater.
Comparing the maximum gas gain measured with all \arch{} mixtures to \arcois{90-10} of similar pressures, $G^\text{max}$ was greater by a factor ranging from $3.2 \pm 0.1$ to $16.2 \pm 0.8$, and on average by a factor of $10.2 \pm 0.7$.
This difference is as expected considering the difference in quencher fraction and the difference in gain expected with \arch{} compared to \arco{}, considering simulations \cite{Deisting_2022,Biagi1018382} and measurements \cite{Sharma_Sauli}.

\subsection{Extrapolation to pressure needed for a long-baseline experiment's ND}
\label{sec:results:subsec:ndgarpressure}

\begin{figure}
\centering
\subfloat[\arcois{90-10}]{\label{sec:results:fig:gainProjections:arco2:vAatG5000}\includegraphics[height=0.25\textheight,trim = 5 5 45 55, clip=true]{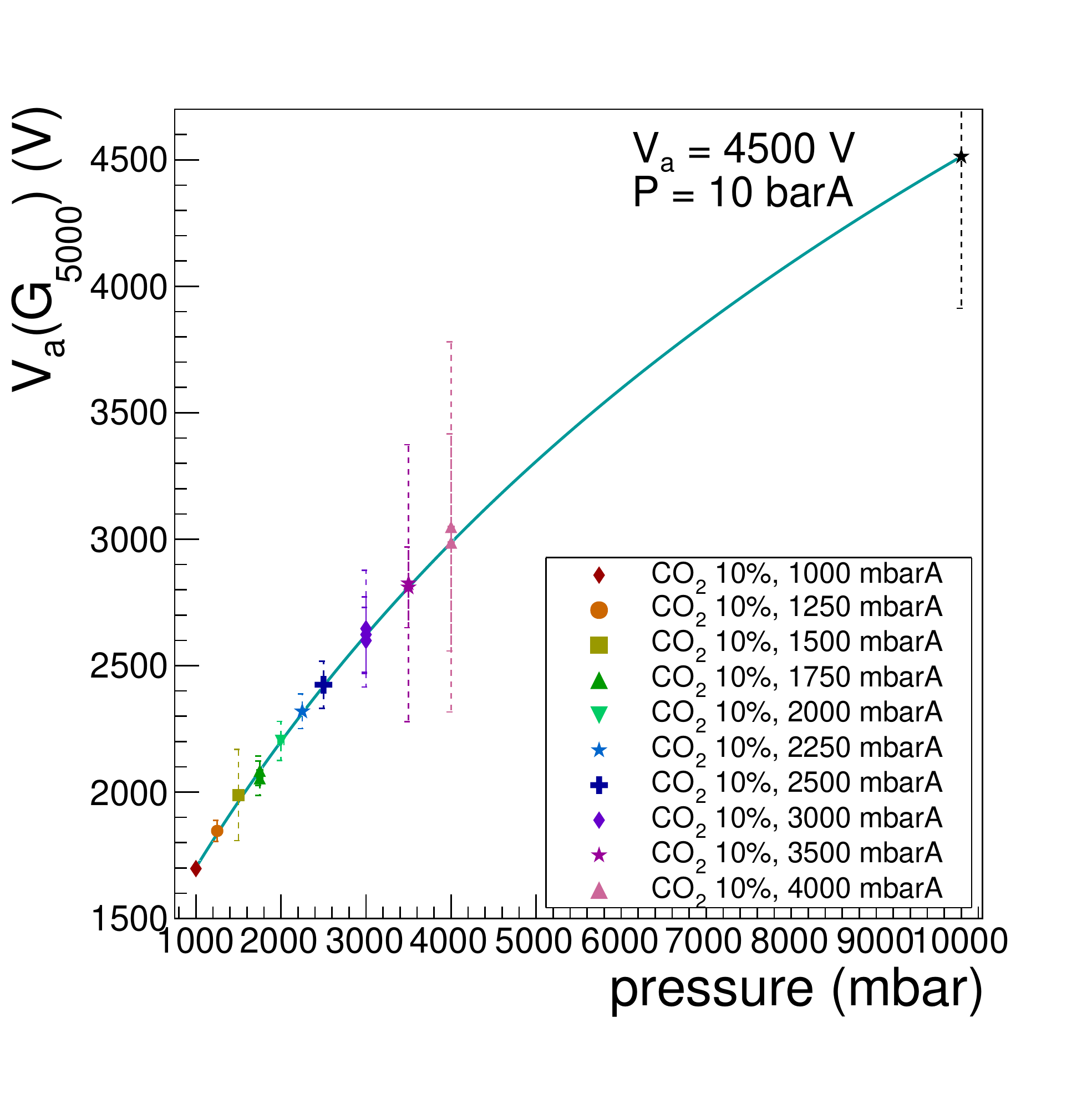}}
\subfloat[\arch{}]{\label{sec:results:fig:gainProjections:arch4:vAatG10000}\includegraphics[height=0.25\textheight]{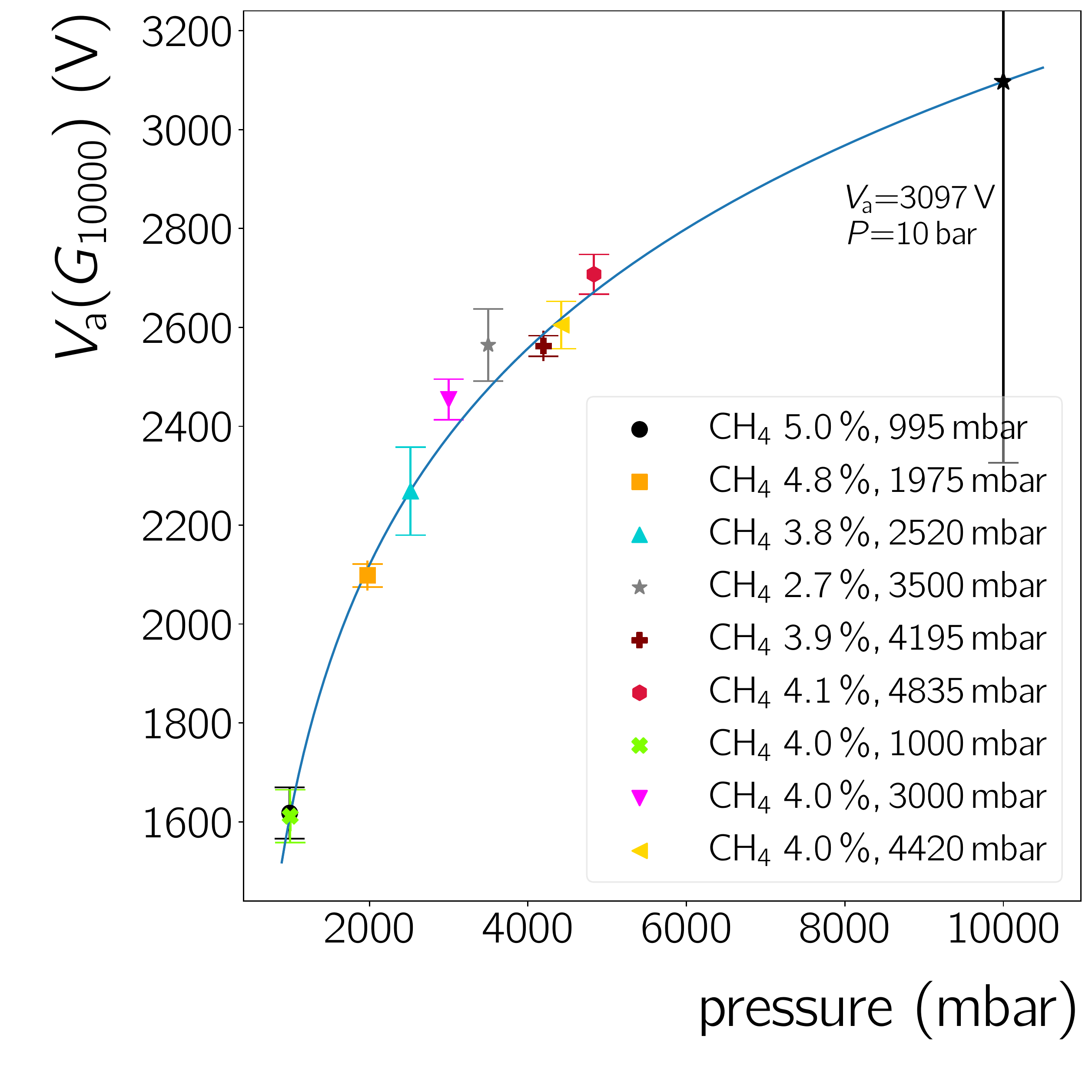}}\\
\subfloat[\arcois{90-10}]{\label{sec:results:fig:gainProjections:arco2:vAatG5000Diethorn}\includegraphics[height=0.25\textheight,trim = 5 5 45 55, clip=true]{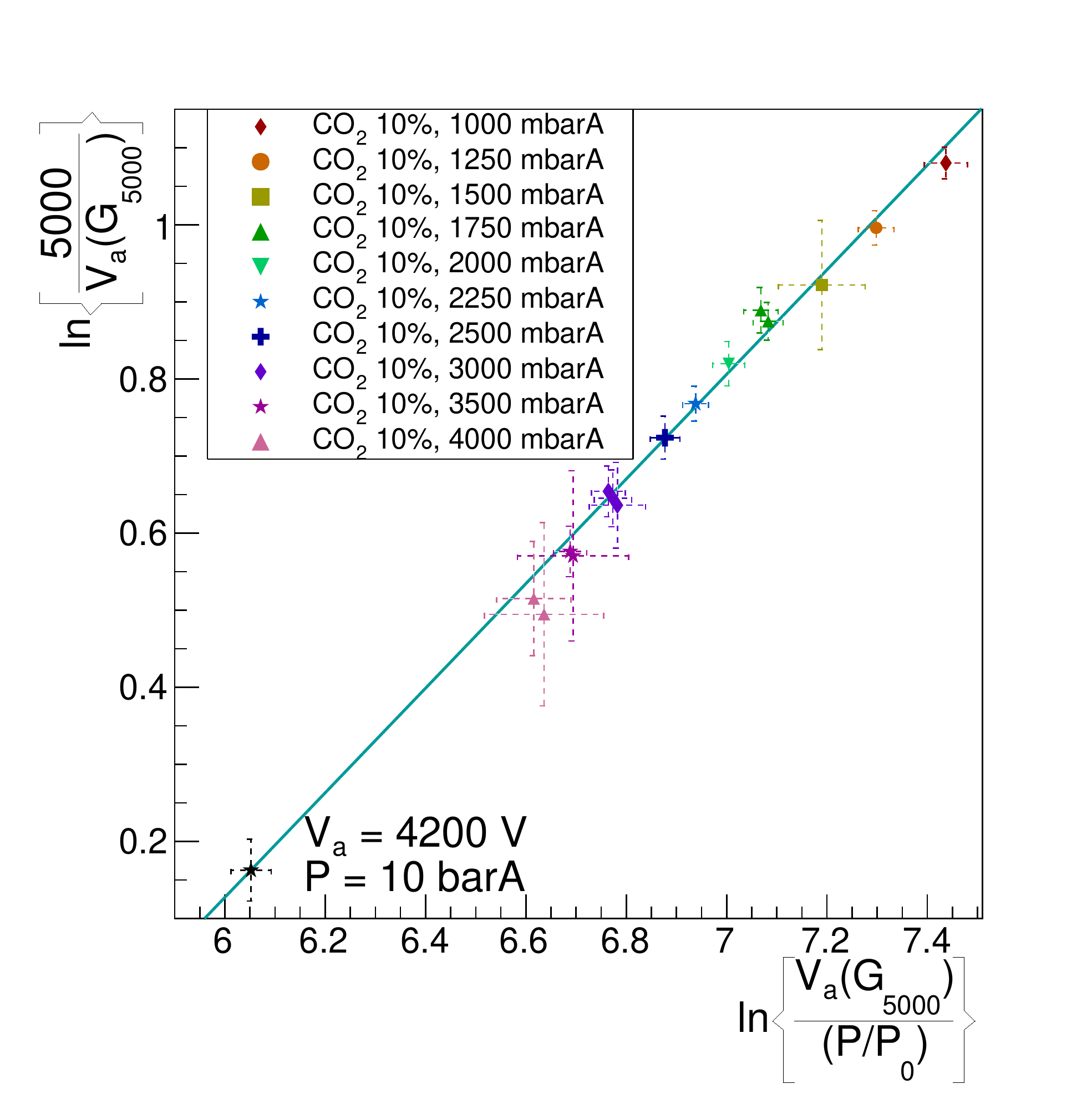}}
\subfloat[\arch{}]{\label{sec:results:fig:gainProjections:arch4:vAatG10000Diethorn}\includegraphics[height=0.25\textheight]{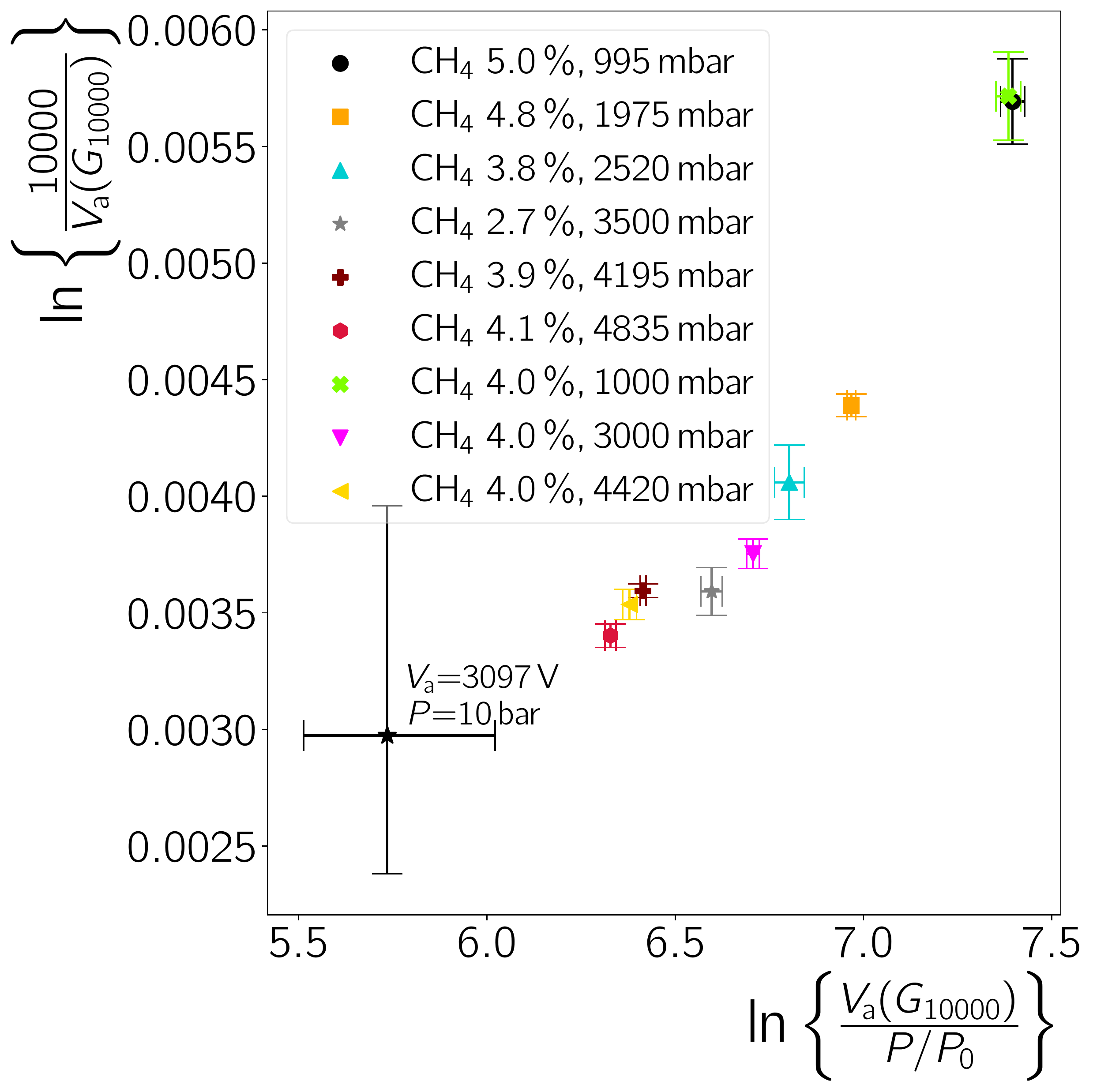}}
\caption{\label{sec:results:fig:gainProjections}Extrapolation of the results from this paper to the tonne-scale detector design pressure of \SI{10}{barA}. The first row plots $V_{\text{a}}(G_{5000})$ ($V_{\text{a}}(G_{10000})$) $\textit{i.e.}$ the voltage at which we measure a gain of 5000 (10,000), against pressure for data taken with \protect\subref{sec:results:fig:gainProjections:arco2:vAatG5000} \arcois{90-10} and \protect\subref{sec:results:fig:gainProjections:arch4:vAatG10000} \arch{} mixtures. 
In each plot, a fit to the data following Eq.~\eqref{sec:results:subsec:ndgarpressure:eq:gainextrapolate} extrapolates to pressures up to \SI{10}{barA}. 
The second row plots the Diethorn relation (Eq.~\ref{sec:results:eq:diethorn}) for $(G_{5000})$ ($(G_{10000})$) for the same data, \protect\subref{sec:results:fig:gainProjections:arco2:vAatG5000Diethorn} \arcois{90-10} and \protect\subref{sec:results:fig:gainProjections:arch4:vAatG10000Diethorn} \arch{}.
The \arcois{90-10} data is fitted with a linear fit and extrapolated to predict a value for $V_{\text{a}}(G_{5000})$ at 10 barA of $\SI{4200\pm600}{\volt}$.
Since several different mixing ratios were used, the Diethorn plot for the \arch{} data is not fitted. Instead, the extrapolated point at \SI{10}{barA} from \protect\subref{sec:results:fig:gainProjections:arch4:vAatG10000} is included in \protect\subref{sec:results:fig:gainProjections:arch4:vAatG10000}.
See \tabref{sec:results:tab:measuredSettings} for the uncertainties on pressure values and mixing ratios.
}
\end{figure}
As expected, the measured gas gain increases exponentially with increasing anode voltage. The lines in the plots in \figref{sec:results:fig:GainVsVoltage_AllPressures} are fits of the exponential function 
\begin{align}
G_\text{OROC}\left(V_{\text{a}}\right) = \text{e}^{m^{\text{gain}}_{\text{exp}}\left(V_{\text{a}} - b^{\text{gain}}_{\text{exp}}\right)}
\label{sec:results:eq:gainparamfit}
\end{align}
Many settings during the \arch{} measurement campaign have been taken where the \fe{55} x-ray photopeak was below the noise threshold in the \textit{Amplitude} histogram. In this case the chosen fit function, Eq.~\eqref{sec:analysis:eq:arch4fit}, returns the position of the noise peak. As these peak positions do not follow the trend expected for exponential avalanche multiplication, all points below $G_\text{OROC}=6000$ have been excluded from the respective fit. 
For both gas mixtures, the slope of the fits ($m^{\text{gain}}_{\text{exp}}$) decreases with increasing gas pressure. 
That is to say that in higher pressure gas mixtures we measure the gas gain to increase more slowly with increasing anode voltage.\\
{\indent}A tonne-scale gas detector with the volume of the ALICE TPC would need to operate at \SI{10}{barA} pressure. By extrapolating $m^{\text{gain}}_{\text{exp}}$ and $b^{\text{gain}}_{\text{exp}}$ (the slope and shift) from the fits in \figref{sec:results:fig:GainVsVoltage_AllPressures} to \SI{10}{barA} pressure, we can comment on the OROC's suitability for operation as a long-baseline neutrino experiment ND with the gas mixtures examined in this paper.
From the Eq.~\ref{sec:results:eq:gainparamfit} fits to the \arcois{90-10} data (\arch{} data), the anode voltage $V_\text{a}$ at which $G_\text{OROC} = 5000$ ($G_\text{OROC} = 10000$), denoted as $G_{5000}$ ($G_{10000}$), is interpolated. 
We deem these to be the sufficient values of $G_{OROC}$ necessary to take a set of measurements like those presented in \secref{sec:analysis}.
\Figref{sec:results:fig:gainProjections:arco2:vAatG5000} and \figref{sec:results:fig:gainProjections:arch4:vAatG10000} plot this voltage, $V_{\text{a}}(G_{5000})$ ($V_{\text{a}}(G_{10000})$), against pressure ($P$) for \arcois{90-10} data (\arch{} data). 
These plots are fitted with the function 
\begin{align}
V_\text{a}\left(G_i\right) = p_0 \cdot \text{ln}\left\{\frac{P-p_1}{p_2}\right\},
\label{sec:results:subsec:ndgarpressure:eq:gainextrapolate}
\end{align}
where $i$ stands for 5000 or 10000 for \arcois{90-10} and \arch{} data respectively. The fit is extrapolated to determine $V_{\text{a}}(G_i)$ for higher pressures up to \SI{10}{barA}. This extrapolation is less motivated for the \arch{} data than it is for the \arcois{90-10}, given that different mixing ratios are used for the first case. As these are mostly in the vicinity of $\sim\!\SI{4}{\%}$ \ch{}, the approach detailed here should still yield a rough estimate for the $P=\SI{10}{barA}$ performance with such a mixture. 
Diethorn's formula \cite{blum_riegler_rolandi} states that 
\begin{align}
\text{ln}\left\{\frac{G_\text{OROC}}{V_\text{a}}\right\} \propto \text{ln}\left\{\frac{V_\text{a}}{P/P_{0}}\right\} \quad.
\label{sec:results:eq:diethorn}
\end{align}
Choosing $P_{0} = \SI{1000}{\milli barA}$, \figref{sec:results:fig:gainProjections:arco2:vAatG5000Diethorn} shows this linear relation for the measured \arcois{90-10} data as well as for the $V_\text{a}\left(G_{5000}\right)$ point extrapolated to $P=\SI{10}{barA}$. In case of \arch{} the relation Eq.~\eqref{sec:results:eq:diethorn} holds too (\figrefbra{sec:results:fig:gainProjections:arch4:vAatG10000Diethorn}), albeit the measured $P=\SI{1}{barA}$ points do not line up well with the other measured points. Due to its large uncertainty the extrapolated point ($V_\text{a}\left(G_{10000}\right)$, $P=\SI{10}{barA}$) fits the other points' distribution. Given that all points should be for the same mixing ratio in order to obey the Diethorn formula, the overall distribution of points seems acceptable.\\
{\indent}For the target pressure of \SI{10}{barA}, the extrapolations in \figref{sec:results:fig:gainProjections} predict $V_{\text{a}}(G_{5000}) = \SI{4500\pm700}{\volt}$ with \arcois{90-10} and $V_{\text{a}}(G_{10000}) = \SI{3100(800)}{\volt}$ with similar \arch{} mixtures.
The additional linear fit to the points in \figref{sec:results:fig:gainProjections:arco2:vAatG5000Diethorn} predicts $V_{\text{a}}(G_{5000}) = \SI{4200\pm600}{\volt}$ at the target pressure of \SI{10}{barA} with \arcois{90-10}. 
These predicted values of $V_{\text{a}}$, which would be necessary to achieve a reasonable gain at \SI{10}{barA} in the stated gas mixtures, are above the range of voltages at which the OROC has currently been tested, so an operation of the OROC at \SI{10}{barA} may not be possible at a sufficient gas gain in either \arcois{90-10} or \arch{} with $\sim\!\!\SI{4}{\%}$ $\text{CH}_4$ content. 
However, the extrapolated \arch{} setting especially is close to the tested range of operation, and a measurement with only \archis{96-4} mixtures may yield that this mixture can reach a sufficient gain at \SI{10}{barA}.
Studies to determine a higher upper limit for $V_\text{a}$ may be necessary; if OROCs can be safely operated at a $V_{\text{a}}$ larger than \SI{3000}{\volt}, then a sufficient gain may be easily achieved at \SI{10}{barA} with these mixtures.

\section{Summary and conclusion}
\label{sec:summary}

A high pressure gas TPC is a good candidate to measure neutrino interactions for future long-baseline neutrino experiments such as the DUNE and HK experiments currently under construction in the United States and Japan, respectively. The ALICE detector has been recently upgraded \cite{ALME2010316} so the existing MWPCs from its pre-upgrade TPC are available as potential readout chambers for such a detector. These MWPCs were however only operated at atmospheric pressure.\\
{\indent}The results presented here demonstrate for the first time the operation of an ALICE outer readout chamber (OROC) at pressures up to \SI{4.8}{barA} in \arch{} mixtures with a \ch{} content between \SI{2.8}{\%} and \SI{5.0}{\%}, and up to \SI{4}{barA} in \arcois{90-10}.
In \SI{4.8}{barA} \archis{95.9-4.1}, a maximum gain of $(64\pm 2)\cdot10^{3}$ was achieved with an anode wire voltage of \SI{2990}{\kilo\volt} and in \SI{4}{bar} \arcois{90-10} a gain of $(4.2\pm0.1)\cdot10^{3}$ was observed with an anode voltage of \SI{2975}{\volt}.\\
{\indent} To achieve a tonne-scale neutrino target in a TPC of the volume of ALICE, a pressure of \SI{10}{barA} would be required. We extrapolate that the OROC anode voltages required to make reasonable gain measurements at this target pressure are $V_{\text{a}} = \SI{4200\pm600}{\volt}$ with \arcois{90-10} and $V_{\text{a}} = \SI{3100(800)}{\volt}$ with an \arch{} mixture with $\sim\!\SI{4}{\%}$ \ch{}. The latter value is less well extrapolated, as not all \arch{} mixtures in the calculation had the same \ch{} content. The Diethorn relation (Eq.~\eqref{sec:results:eq:diethorn}) states that the relationship between $\text{ln}\{\frac{G_\text{OROC}}{V_\text{a}}\}$ and $\text{ln}\{\frac{V_\text{a}}{P/P_{0}}\}$ should be linear and our measurements as well as the extrapolated points are consistent with this relationship.\\
{\indent}The so identified $V_{\text{a}}$ for operation at \SI{10}{barA} are above the maximal anode voltage at which the OROC has been tested, so an operation of the OROC at \SI{10}{barA} may not be possible with sufficient gas gain in either \arcois{90-10} or \arch{} with $\sim\!\!\SI{4}{\%}$ $\text{CH}_4$ content. 
Studies to determine a higher upper limit for $V_{\text{a}}$ may be necessary. If $V_{\text{a}}$ larger than \SI{3000}{\volt} would be deemed to be safe for the wire chambers then, in \arch{} in particular, it may be possible to achieve higher pressures at reasonable gain values with the gas mixtures tested in this paper. The maximal $V_{\text{a}}$ achievable without damage to the readout chambers has not been investigated and the limit of \SI{3}{\kilo\volt} used here is imposed as a precaution to protect the equipment while the maximum safe $V_{\text{a}}$ is still unknown. 
The maximum pressure was limited by the rating of the UK high pressure plattform \cite{instruments5020022}, which can be operated only at pressures up to \SI{5}{barA}. Other groups in the US are carrying out tests with a smaller inner ROC of the ALICE TPC at test-stand \cite{tanaz2022} that will be able to reach 10 barA.\\
{\indent}Several optimisations could improve our results.
The energy resolution for these measurements is generally between \SI{10}{\%} and \SI{20}{\%}, but we obtained values as high as \SI{60}{\%} for the highest pressure \arch{} measurements. This could be improved with better noise reduction and with more precise positioning of the \fe{55} source in the vessel, neither of which were optimised. The energy resolution can likely also be improved with better background subtraction at the analysis stage.
Optimizations to improve energy resolution will become increasingly important for future studies, which are in preparation right now. Currently custom readout electronics based on the SAMPA chip are close to being finalised and will be tested in the near future; single pad readout and charge pre-amplification close to the pads will not only yield a better energy resolution but also allow for particle tracking.

\section*{Acknowledgements}

The authors would like to gratefully acknowledge the help of Ian Murray with sourcing lab equipment, gas orders and general support during the operation of the detector, and the help of Richard Elsom, Paul Bamford and Ian Higgs with the production of essential detector parts. We further thank Chilo Garabatos on behalf of the ALICE Collaboration for sending the OROC used in this paper to the UK. This project has received funding from the European Union’s Horizon 2020 Research and Innovation programme under grant agreement No 101004761 and an EU Horizon 2020 Marie Skłodowska-Curie Action. This research was furthermore funded in part by Science and Technology Facilities Council grant number ST/N003233/. We thank UKRI for supporting several of the authors of this paper.

\AtNextBibliography{\small}
\sloppy
\printbibliography[heading=bibintoc]

\end{document}